\documentclass[13pt]{article}
\usepackage{epstopdf}
\usepackage{cite}
\usepackage{authblk}
\usepackage[top=30truemm,bottom=30truemm,left=25truemm,right=25truemm]{geometry}
\usepackage{graphicx}% Include figure files
\usepackage{bm}% bold math
\begin{document}

%\preprint{}

\title{
Gliding filament system giving both orientational order and clusters in collective motion
}


\author{Sakurako Tanida\textit{$^{1}$}, 
Ken'ya Furuta\textit{$^{2}$}, 
Kaori Nishikawa\textit{$^{1}$}, 
Tetsuya Hiraiwa\textit{$^{1}$}, 
Hiroaki Kojima\textit{$^{2}$},
Kazuhiro Oiwa\textit{$^{2}$}, and
Masaki Sano\textit{$^{1}$}
}
%\author{Sakurako Tanida\textit{$^{1}$}, 
%Ken'ya Furuta\textit{$^{2}$}, 
%Kaori Nishikawa\textit{$^{1}$}, 
%Tetsuya Hiraiwa\textit{$^{1}$}, 
%Hiroaki Kojima\textit{$^{2}$},
%Kazuhiro Oiwa\textit{$^{2}$}, and
%Masaki Sano\textit{$^{1}$}
%\thanks{}
%\affiliation{
%$^{1}$Department of Physics, Universal Biology Institute, Graduate School of Science, the University of Tokyo, 7-3-1 Hongo, Bunkyo-ku, Tokyo, Japan\\
%$^{2}$National Institute of Information and Communications Technology, 588-2 Iwaoka, Iwaoka-cho, Nishi-ku, Kobe, Japan
%}

\affil{$^{1}$Department of Physics, Universal Biology Institute, Graduate School of Science, the University of Tokyo, 7-3-1 Hongo, Bunkyo-ku, Tokyo, Japan\\$^{2}$National Institute of Information and Communications Technology, 588-2 Iwaoka, Iwaoka-cho, Nishi-ku, Kobe, Japan}

\date{\today}
\maketitle
\begin{abstract}
Active matter consists of self-propelled elements exhibits fascinating collective motions ranging from biological to artificial systems. Among wide varieties of active matter systems, reconstituted bio-filaments moving on molecular motor turf interacting purely by physical interactions provides the fundamental test ground for understanding biological motility. However, for the emergence of ordered patterns such as polar pattern, swirls, clusters, and density wave in actomyosin motility assay, multi-filament collisions are required instead of binary collision which is often assumed in kinetic theory. Similarly, for microtubules driven by kinesin motors to produce nematic ordered state, depletion agents or binding molecules are required to introduce strong alignment effects between filaments. Thus, whether simple physical interactions during collisions such as steric effect without depletion nor binding agents are sufficient or not for producing ordered patterns in motility assays remains still elusive. In this article, we constructed a motility assay purely consists of kinesin motor and microtubule in which the frequency of binary collision can be controlled without using depletion nor binding agents. By controlling strength of steric interaction and density of microtubules, we found different states; disordered state, long-range orientationally ordered state, liquid-gas-like phase separated state, and transitions between them. We found that a balance between cross over and aligning events in collisions controls transition from disorder to global ordered state, while excessively strong steric effect leads to the phase separated clusters. Furthermore, macroscopic chiral symmetry breaking observed as a global rotation of nematic order observed in this experiment could be attributed to the chirality at molecular level.
Numerical simulations in which we change strength of volume exclusion reproduce these experimental results. Moreover, it reveals the transition from long-range alignment to nematic bands then to aggregations. 
This study may provide new insights into dynamic ordering by self-propelled elements through a purely physical interaction.
\end{abstract}
%\pacs{}
%\keywords{}

%
\section{Introduction}
Collective behavior of motile elements is ubiquitous in wide varieties of living systems ranging from molecular level to cellular level and even in individual animal levels. 
In these hierarchical systems, each motile element dissipates energy and transduces it into motion \cite{Vicsek2012,Wu2009,Zhang2010,Peruani2012,Nishiguchi2017,Kawaguchi2017,Szabo2006,John2009a,Tennenbaum2015,Gerum2013,Lopez2012,Katz2011,Ballerini2008a,Helbing2000}.
The group of such objects can organize sustained collective motion at each level as in cytoplasmic streaming driven by molecular motors, moving clusters in bacterial colonies, swirling in cell tissues, and flocking of birds. 
Interactions in these systems diversely include mechanical, chemical, and informational processes. 
Statistical mechanics manifests that details of the elements and interactions become irrelevant for a large system size limit, and global behavior is determined by a few factors such as dimensionality and symmetries of the systems thanks to the universality classes. 
This is the common belief that collectives of motile elements can be regarded theoretically as active matter systems irrespective of whether the system is living or artificial {\cite{Vicsek1995,Toner1995,Simha2002,Gregoire2004,Kraikivski2006,Ginelli2010,Yang2010,Abkenar2013,Narayan2007a,Bricard2013,Nishiguchi2018}. 
Nevertheless, subtle difference in interactions sometimes makes all the difference. 
Such an example exists even in the fundamental motile systems consist of molecular motors and bio-filaments which cause most of complex motile behaviors in living systems. 
In a large parameter space of these systems, diverse patterns are observed. Despite great research efforts, general understanding of what difference in interactions of constituents bring different patterns still remains elusive. 
In this article, we constructed a motility assay consists of kinesin motor and microtubule in which the frequency of binary collision can be controlled without using depletion nor binding agents. 
By controlling strength of volume exclusion interaction and density of filaments, we found different states; disordered state, long-range orientational ordered (LOO) state, liquid-gas-like phase separated (LGS) state, and transitions between them. 
We found that a balance between crossing-over and aligning events between colliding filaments controls transition from disordered to LOO states, while excessively strong volume exclusion interaction leads to a cluster (LGS) state. 
This finding gives a unified view for seemingly different phases; cluster phase and long range nematically ordered phase in suspension of bacteria observed in independent experiments, as two limiting cases expected to appear by varying excluded volume interaction\cite{Zhang2010,Peruani2012,Nishiguchi2017}.

The reconstituted cytoskeletal systems consists of the smallest elements at the size of nano meter scale, thus they have a possibility to attain the maximum system size (the ratio of the system size to elementary size) among a variety of active matter systems. 
There are several combinations in these systems.
The first combination is actin filaments and myosin motors. High concentration of actin filaments led to polar pattern \cite{Schaller2010,Butt2010,Hussain2013a}, loops \cite{Schaller2011}, swirls, clusters, and density wave \cite{Schaller2011a,Schaller2013,Suzuki2017}, where emergence of ordered patterns could not be explained solely from alignment by binary collision of filaments, instead multi-filament collisions are required\cite{Suzuki2017}. The second combination is microtubule (MT) and dynein motors \cite{Sumino2012}. In this case, the ratio of crossing-over to all events was about 30\% in binary collisions. It suffices to give rise to an ordered pattern, however, MTs eventually formed large ``vortices" in which MTs aligned their orientations nematically with each other. 
Origin of vortex pattern was attributed to the intrinsic curvature of MTs trajectories and its memory effect. 
The third combination is MT and kinesin motors,in which following several patterns were known to emerge \cite{HenryHess2005,Kawamura2008,Tamura2011,Liu2011b,Kabir2012,Inoue2013,Lam2014a,Inoue2015,Saito2017}. One is tiny loop arising from the steric barrier between MT segments \cite{Kawamura2008,Tamura2011,Liu2011b,Kabir2012,Inoue2013}.
Another pattern is a ``stream", which shows global alignment appearing at high MT density \cite{Inoue2015,Saito2017}. 
In order to realize global order, either depletion agents (methylcellulose) or binding molecules for MTs were required\cite{Kakugo2011}. 
Also, asters and topological defects appear in MT and kinesin mixed solution \cite{Surrey2001,DeCamp2015,Torisawa2016a}.

In this paper, we focus on the reconstructed system of MTs and kinesins.
When kinesin motors are immobilized on the substrate with non-specific binding, no significant pattern has emerged in motility assay composed of only MTs and kinesins in consistent with previous reports.
The fact would be the consequence that a large part of kinesins are not functioning due to non-specific binding to glass surface and hight of functioning kinesins were not uniform, which results in lack of collision interaction between MTs. 
To solve this problem, we coated glass surface with surfactant Pluronic F-127 to avoid non-specific binding. 
F-127 is tri-block copolymer of ABA type with A being hydrophilic and B being hydrophobic. We introduced a specific binding spot for kinesin into the two footing parts A which stem from the central part B on the substrate.
The height of Pluronic F-127 molecule is about 30 nm, the size of kinesin motors bound to F-127 is about 10 nm, and the diameter of MT gliding on kinesins is 25 nm.
This procedure to make the height of kinesin even, microtubules can be confined into two dimensions and then MTs always collide and align without overlapping when all F-127 molecules are functionalized. 

Owing to treatment to promote collision interaction, we found dense gliding MTs form ``liquid-gas-like phase separation(LGS)'' pattern.
This LGS pattern is rather similar to those in bacteria gliding on an ager surface exhibiting moving cluster pattern\cite{Wu2009,Zhang2010,Peruani2012}.
Considering that bacteria swimming highly confined to a quasi-2-dimensional thin chamber showed global alignment pattern \cite{Nishiguchi2017}, overlapping of moving particles seemed to enable them to generate non-cluster pattern in two dimensions.
Also, numerical simulations of motile rods changing the strength of volume exclusion have shown the formation of moving clusters and lanes \cite{Abkenar2013}.
Those suggested the possibility that MTs form other patterns if we could control the ratio between aligning and cross over events during collision.

To do this, we decreased kinesin density by reducing the ratio of concentration of functionalized Pluronic to that of non-functionalized one, kinesins can bind only the sites of functionalized Pluronic.
At the limit of high kinesin density, MTs are forbidden to overlap because of tight adhesion to kinesin-coated surface at uniform height promoting collisions (Fig. \ref{fig:sketch}, right).
In contrast, at the lower kinesin density, MTs can overlap because intervals of kinesins attaching on a single MT are wide and the tip of the MT is free from the confinement of kinesin-coated surface (Fig. \ref{fig:sketch}, left).
With this lowered kinesin density assay, we found that gliding MTs form another new pattern, ``long-range orientational ordered (LOO)", which is similar to a global alignment pattern of swimming bacteria highly confined to a quasi-2-dimensional thin chamber \cite{Nishiguchi2017}.
Since, MTs are about 5 $\mu$m length and those in this pattern were aligning at least an 1 mm $\times$ 1 mm region.
It is the largest system showing global alignment. 

Whether objects can overlap or not is interpreted as a degree of strength of volume exclusion.
Stronger volume exclusion can align pair of colliding MTs more frequently, so that it might be thought to form global alignment.
Surprisingly, our results suggest that weak volume exclusion forms LOO state while strong volume exclusion forms LGS.
In addition, as shown in Fig.\ref{fig:global_plots}, we found that the phase transition occurred depending on MT density.
With weak volume exclusion, {\it i.e.} at the lower kinesin density, increasing the MT density led to the the phase transition from disordered to LOO patterns through phase coexistence regime.
In contrast, with strong volume exclusion, {\it i.e.} at high kinesin density, it led to phase transition from disordered to LGS patterns.
Moreover, to confirm this tendency, we carried out numerical simulations applying the particle-based
model of self-propelled objects with volume exclusion. Numerical results followed our experimental observations.
Consequently, these results suggest that the above-mentioned two patterns observed in a system with unidirectional motion and nematic alignment could be explained by the parameter of volume exclusion.
Also we report striking behaviors, global rotation of orientation at low kinesin density and loop formation at high kinesin density.

%________________________________________________________________________________
\begin{figure}[t]
\centering
\includegraphics[width=.5\linewidth]{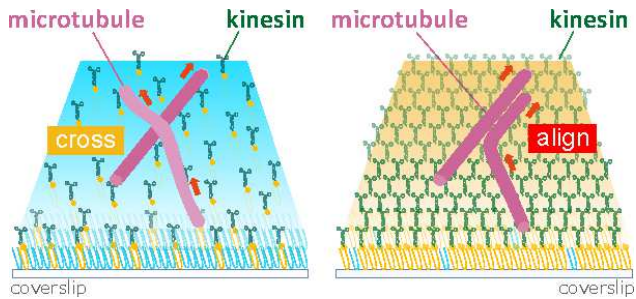}
\caption{Schematic images of the motility assay controlling kinesin density. When the kinesin density is low, the intervals of the kinesins attaching on a MT are wide enough for MTs to crossing over (left). When it is high, MTs adhere to glass surface through kinesin, preventing them from crossing over (right).}
\label{fig:sketch}
\end{figure}
%___________________________________________________________________________________

The organization of this paper is as follows: Section \ref{sec:MateMeth} is devoted to brief descriptions of experimental preparation and method.
In Section \ref{sec:strength}, we examine kinesin density can control the strength of volume exclusion in our experimental system.
We analyze the effect of volume exclusion on global pattern of both experiment and numerical simulation in Section \ref{sec:global}.
To explain the global behavior, we study binary collision interaction in Section \ref{sec:binarycollision}.
We discuss and then summarize our results in Section \ref{sec:discussion} and \ref{sec:conclusions}.

\section{Materials and Methods}
\label{sec:MateMeth}

\subsection{Preparation of MT and kinesin}
Tubulins were purified from porcine brain using a high-molarity PIPES buffer (1 M PIPES, 20 mM EGTA, and 10 mM MgCl$_2$; pH adjusted to 6.8 using KOH) as described previously  \cite{Castoldi2003}.
Rhodamine-labeled tubulins and Alexa 488-labeled tubulins were prepared using 5-(and-6)carboxy-tetramethyl-rhodamine succinimidyl ester (Invitrogen, C1171), and Alexa Fluor 488 succinimidyl ester (Alexa Fluor 488-SE\textregistered; Invitrogen A20000), respectively, according to the standard techniques \cite{Hyman1991}.
These two differently labeled tubulins ($8\ \%$ labeled, $3.0$ mg/ml) were separately polymerized into MTs in the presence of guanosine-5'-[($\alpha,\beta$)-methyleno]triphosphate (GMPCPP) ($1$ mM), and BRB80 buffer (80 mM PIPES-KOH, 1 mM MgSO$_4$, 1 mM EGTA, pH 6.8) in $37\  ^\circ $C for $30$ min. 
The MT solution was diluted to $0.15 \ {\rm mg/ml} $ with taxol (final $50 \ {\rm \mu M}$) and left for at least 2 days.
Length of MTs prepared in this method ranged from 4 to 8$\  {\rm \mu m}$ (See Supplementary Information (SI) \cite{SI}).
The final solution was made by mixing these two solutions labeled with different fluorophores  such MTs that the ratio of the number of Alexa 488-labeled MTs to the total MT number ranged between 1 and $ 5\  \%$.
The SNAP-tagged kinesin-1 (rat KIF5C truncated at 430 amino acids, and fused with SNAP-tag and six-histidine tag at the C-terminus) was expressed in {\it Escherichia coli} Rosetta2 (DE3) and prepared as described previously \cite{Furuta2013}.

\subsection{Synthesis of BG-functionalized Pluronic F-127}
The terminal hydroxyl groups of Pluronic F-127 were functionalized with benzylguanine via an amine terminated Pluronic F-127 \cite{Li1996}. Pluronic F-127 (2 g, 0.16 mmol; Sigma-Aldrich, P2443) was dissolved in 6 ml of benzene and slowly added into a stirred solution of 4-nitrophenyl chloroformate (192 mg, 0.954 mmol; Tokyo Chemical Industry, C1400) in 6 ml benzene. The solution (12 ml) was continuously stirred at room temperature for 24 h under nitrogen atmosphere. The resulting yellow-colored solution was mixed with 200 ml of ice cold diethyl ether for 5 min using a magnetic stirrer. The precipitate was then collected by vacuum filtration (Millipore, model WP6110060; TGK, 0371430103) through a membrane filter with 47 mm in diameter (Whatman, 1450-090; manually cut with a circle cutter). The precipitate was further washed by 4--5 additional cycles of dissolving and precipitating until the precipitate turns white and then dried under vacuum overnight. 
 The activated F-127 ($\approx$1 g) was dissolved in methanol. One milliliter of hydrazine monohydrate solution (Wako, 081-00893) was added drop-wise and stirred at room temperature for 16 h under nitrogen atmosphere. The product was precipitated with 120 ml of diethyl ether and vacuum filtered (Sansyo, 81-0106) through a 90-mm membrane filter. The precipitate was washed four times.
 The vacuum dried F-127-amine (6.4 mg) and an amine-reactive benzylguanine reagent (2 mg; New England Biolabs, BG-GLA-NHS, S9151S) was dissolved in 0.3 ml of anhydrous {\it N,N}-dimethylformamide, allowed to stand at room temperature overnight, and dried under vacuum for at least 4 hours. The product was then dissolved in 0.65 ml of Milli-Q water. The precipitated benzylguanine reagent was removed with a spin filter (Millipore, UltraFree 0.1 $\mu$m, UFC30VV25). The filtrate was further filtered through a Zeba spin filter (Thermo Scientific, 7k MWCO, 89882) to remove dissolved benzylguanine reagent. The filtrate was stored at $-$80${}^\circ$C.

\subsection{Preparation of flow chamber}
Teflon-coated coverslips were prepared as described previously \cite{Furuta2017}.
A flow chamber was made of $22 $ mm $\times 32$ mm Teflon coating coverslip and $18\  {\rm mm} \times 18\  {\rm mm}$ coverslip with parafilm as a spacer. 
The flow chamber was first filled with Pluronic F127 functionalized with benzylguanine.After 10 minutes incubation, the flow chamber was washed out with BRB80 (50 ml) and introduced 0.25 mg/ml SNAP-tagged kinesin-1 solution.
After another 10 minutes incubation and washing, we filled with the dual-colored MT solution described above into the chamber. The flow chamber was again incubated for 5--10 minutes to allow the MTs to bind kinesins, and then washed out with Assay buffer ($10\ \mu$M taxol, 25 mM glucose, and $200\ {\rm\mu g/ml}$ glucose-oxidase, $40\ \mu$g/ml catalase, and $140$ mM bata-melcaptoethanol in BRB80). Finally, we introduced ATP solution with the ATP regenerating system (10 mM ATP, 2 unit/ml pyruvate kinase/lactate dehydrogenase, and 2.5 mM phosphoenol-pyruvate in Assay buffer) into the flow chamber.

To control the kinesin density on the glass surface of a chamber, we changed ratio of Pluronic F127 functionalized with benzylguanine.
The mixing ratios of functionalized Pluronic F127 to non-functionalized one were adjusted to be $5\%$ and $100\%$. Each made kinesin density $(6.1 \pm 0.7) \times 10^3$ molecules per ${\rm \mu m^2}$ and $(2.5 \pm 0.6) \times 10^3$ molecules per ${\rm \mu m^2}$, respectively.

\subsection{Microscopy and image capture}
To observe the motility of MTs, samples were illuminated with a 100 W mercury lamp and visualized by Leica DMI 6000B using a objective lens HCX PL APO 63x/1.40-0.60 OIL CS (Leica). Images were captured using EMCCD camera (iXon Ultra, Andor) connected to a PC. 
For observations of binary collision, snapshots were taken at every 10 sec for 3 hours. 
And for observation of patterns, we scanned and stitched multiple area to get a wide snapshot.

Images and movies of motility assays of MTs captured by the fluorescence microscopy were analyzed using the image analysis software ImageJ, and the algorithm which we have developed in python with scikit-image and python image library.

%%%%%%%%%%%%%%%%%%%%%%%%%%%%%%%%%%%%%%%%%%%%%%%%%%%%%%%%%%%%%%%%%%%%%%%%%%%%%%%%%%%%%%%%%%%%%%%%
\section{Results}
\label{sec:results}
\subsection{Strength of volume exclusion depending on kinesin density}
\label{sec:strength}
%-------------------------
\begin{figure}[tbhp]
\centering
\includegraphics[width=.6\linewidth]{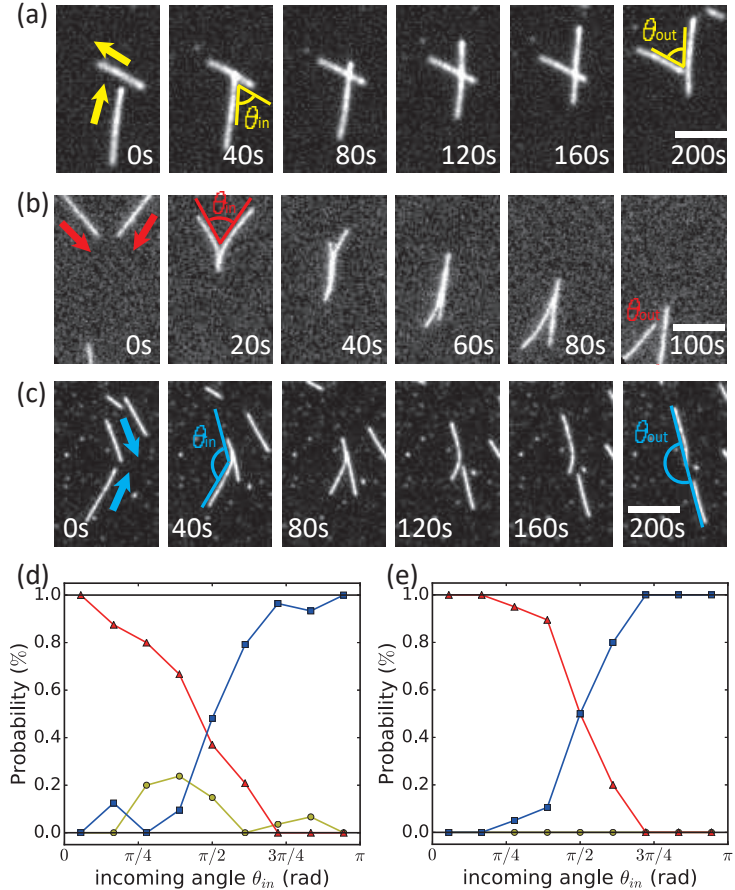}
\caption{
(a-c) Snapshots of three types of behavior after collision: crossing-over (a), alignment (b), anti-alignment (c).  
Scale bar $8\  \mu m$.
(d,e) Probability of each type in each angular bin
at the lower kinesin density (d, number of collision event is $n=147$)  and at the higher kinesin density (e, $n=135$).
Yellow circle markers represent crossing-over; red triangle markers represent alignment;
blue square markers represent anti-alignment.
}
\label{fig:snaps}
\end{figure}
%-------------------------
\clearpage
To assess the effect of kinesin density on the volume exclusion, we observed behaviors of two MTs during collision at each kinesin density.
The experiments were conducted under a dilute condition, $0.4\times 10^{-3}$ filaments per ${\rm \mu m^2}$. 
MTs showed isotropic and homogeneous state at this diluted density. 
The behaviors after collision can be classified three types; crossing-over, alignment, and anti-alignment  [Fig. \ref{fig:snaps} (a)-(c)].
We firstly define crossing-over as collision events having one or more snapshots in which the observing pair of MTs forms four branches [the third to fifth snapshots in Fig. \ref{fig:snaps} (a)]. 
From the rest of events, alignment was defined such that the outgoing angle $\theta_{out}$ is smaller than $\pi/2$, and anti-alignment is defined otherwise. 
Incoming and outgoing angle $\theta_{in}, \theta_{out}$ are defined as the angle at which two MTs touch and detach, respectively.

As mention in Sec \ref{sec:MateMeth}, we prepared and observed chambers at two different kinesin densities.
At the higher kinesin density of $(6.1 \pm 0.7) \times 10^3$ molecules per ${\rm \mu m^2}$, no crossing-over was observed  [Fig.\ref{fig:snaps} (e)].
On the other hand, $10\%$ of events showed crossing over at the lower kinesin density of $(2.5 \pm 0.6) \times 10^3$ molecules per ${\rm \mu m^2}$ [yellow circle marker in Fig.\ref{fig:snaps} (d)].
MTs can crossing-over only when the volume exclusion is weak, so that these results indicate that the strength of the volume exclusion can be controlled via kinesin density.
Although crossing-over ratio for the higher kinesin density is not very different from that for the lower kinesin density, but it is enough to produce different behaviors.
Whether crossing-over is possible or not depends on whether the tip can surmount the height of the MT.
The height of a MT is about 25 nm, and it is almost the same as the average interval of kinesin at the lower kinesin density.
It would generate crossing-over after collision as shown in Fig.\ref{fig:snaps} (d).

In addition to the density-dependent behaviors, Figs. \ref{fig:snaps} (d) and (e) clearly give angle-dependent behavior. Alignments tend to occur at acute incoming angle, $\theta_{in}<\pi/2$, and anti-alignments tend to occur at obtuse incoming angle, $\theta_{in}>\pi/2$ at both kinesin densities [Figs.\ref{fig:snaps} (d) and (e)]. This implies that velocity alignment is nematic in this experimental system.

%%%%%%%%%%%%%%%%%%%%%%%%%%%%%%%%%%%%%%%%%%%%%%%%%%%%%%%%%%%%%%%%%%%%%%%%
\subsection{Global patterns} 
\label{sec:global}

\subsubsection{Appearing patterns depending on kinesin density} 
\label{sec:kinesindensitydep}
%-------------------------
\begin{figure*}[tbph]
	\includegraphics[width=.99\linewidth]{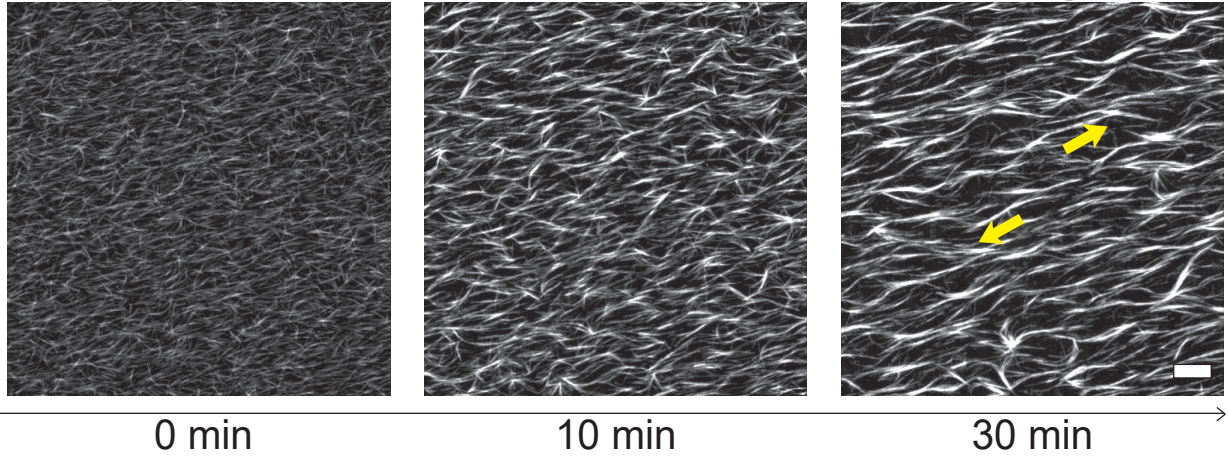}
	\caption{Time evolution of MT patterns at the higher and the lower kinesin densities. Yellow arrows show the directions of motions of representative MTs. At initial state at both kinesin densities, MTs were bound to the kinesin-coated surface randomly. After injection of ATP solution, MTs gradually formed patterns. At the higher kinesin density, MTs gathered and went into a ``liquid-gas-like phase separation  (LGS)" state  without nematic orientational order (a). In contrast at the lower kinesin density, nematic orientational order was increasing with time and pattern ended up with `` Long-range orientational order (LOO)" state (b). The MT densities are 0.3 (a) and  0.1 (b) filaments par ${\rm \mu m^2}$. Bar is 10 ${\rm\mu m}$. }
	\label{fig:global_snaps}
\end{figure*}
%-------------------------
As mentioned above, our experimental set up enabled us to control the strength of volume exclusion (See Sec. \ref{sec:strength}).
Using this set up, we found two different patterns depending on the kinesin density. 
Figures \ref{fig:global_snaps}(a) and (b) show time evolution of patterns at the higher and lower kinesin densities, respectively. At the higher kinesin density, MTs started to segregate spontaneously and go into the ``liquid-gas-like phase separation (LGS)" state forming clusters which were moving around in random directions [Fig. \ref{fig:global_snaps}(a)]. In an individual cluster, MTs were aligned in parallel and almost all of them moved in the same direction, thus forming polar cluster. Crawling around, the clusters often merged with other and sometimes splitting into small polar clusters. 
In contrast at the lower kinesin density, gliding MTs eventually aligned their orientation and went into the ``long-range orientational ordered (LOO)" state in 30 minutes [Fig. \ref{fig:global_snaps}(b)]. In this state, MTs move in both directions.

%________________________________________________________________________________
\begin{figure*}[htbp]
	\centering
	\includegraphics[width = 1.\linewidth]{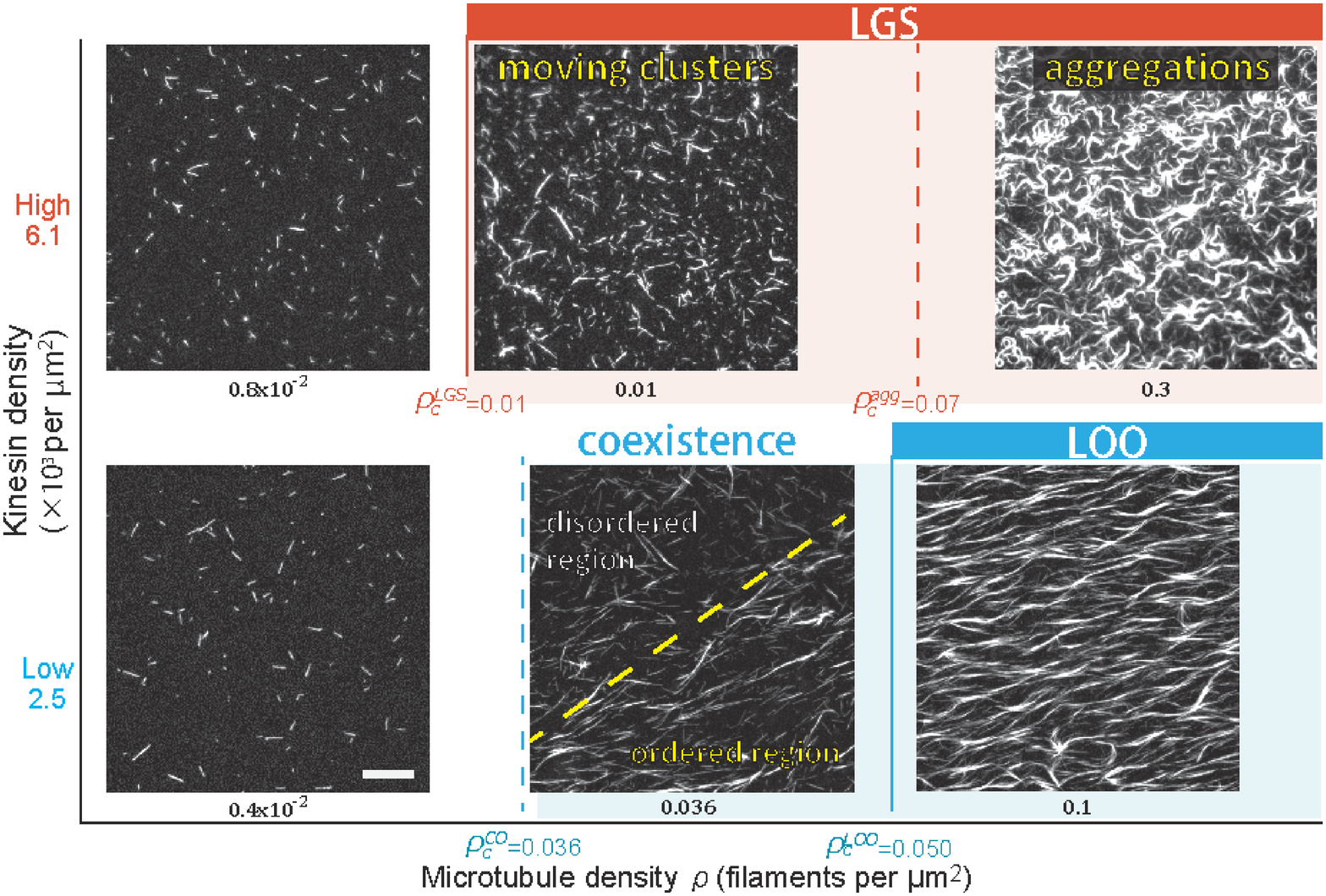}
	\caption{
Phase diagram for various kinesin and MT densities.
At the higher kinesin density, pattern was transited from disordered to order state at 0.01 filaments per ${\rm \mu m^2}$ .
Moreover, ordered state at the higher kinesin density was categorized by dynamics of clusters which emerged in this ordered state.
Clusters were moving around below 0.07 filaments per ${\rm \mu m^2}$(``moving cluster"), while they got into and formed ``aggregation" above the MT density.
At the lower kinesin density, 
when the filament density is lower than 0.032 per ${\rm \mu m^2}$, the disordered state is observed, whereas
when the density is higher than it, the coexistence of ordered and disordered patterns is observed.
The border between ordered and disordered regions is represented by a yellow broken line in a bottom middle image, whose kinesin density is low and MT density is 0.036 filaments per ${\rm \mu m^2}$. Above 0.05 filaments per ${\rm \mu m^2}$, MTs showed the LOO pattern.
Bar is 20 $\mu$m.}
\label{fig:global_plots}
\end{figure*}
%_______________________________________________________________________________________

\subsubsection{Critical densities depending on kinesin density}

%_______________________________________________________________________________________
\begin{figure*}[tbp]
	\centering
	\includegraphics[width=.99\linewidth]{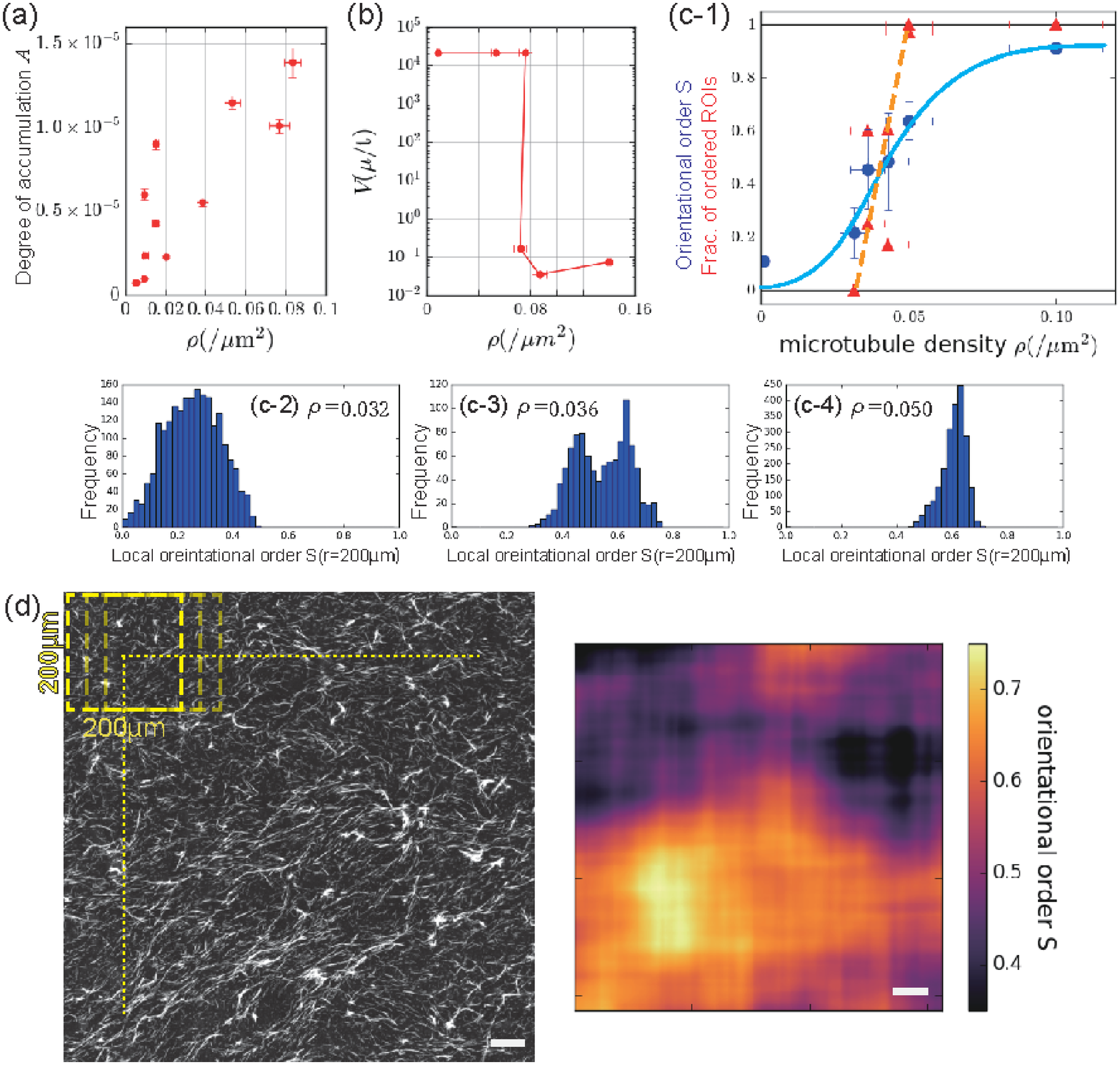}
	\caption{
(a) Degree of accumulation $A$ for various MT densities at the higher kinesin density. It begun to increase at $0.01$ filaments per $\mu m^2$ and form clusters. 
(b) Degree of the moving velocity of structure at the higher kinesin density $V$ for various MT densities. $V$ drops abruptly at $\rho =0.07$ because ``aggregation" of cluster becomes dominant.
(c-1) Nematic orientational order $S$ (blue) and fraction of ordered area (red) for various MT densities at the lower kinesin density. Blue circle markers and vertical error bars represent mean and standard deviation of $S$, respectively.
Blue horizontal error bar is standard deviation of MT densities.
A cyan line is trend line of $S$ for eye guide.
Red triangle markers represent fractions of ordered area which are calculated in local area $200\  {\rm \mu m} \times 200\  {\rm \mu m}$ and $S>0.5$ at high kinesin density.
Red horizontal error bars represents standard deviation of MT densities. An orange broken line shows trend line of  fractions of ordered area for eye guide.
Histograms show distributions of local orientational order for 0.032, 0.036, and 0.05 filaments per ${\rm \mu m^2}$ (c-2,3,4).
At 0.036 filaments per ${\rm \mu m^2}$ (c-3), the distribution is bimodal, which means order-disorder phase coexistence. On the other hand, 0.032 and 0.050 filaments per ${\rm \mu m^2}$ (c-1,2), the distribution is unimodal.
(d)Coexistence sate: A $825.5 \ {\rm \mu m} \times 825.5\  {\rm \mu m}$ snapshot of coexistence of ordered and disordered regions, and color map of spacial distribution of local orientation. Local orientations are calculated in $200\  {\rm \mu m} \times 200\  {\rm \mu m}$ area, which moves to cover the whole image as displayed in left. Bar is 20 $\mu $m
}
	\label{fig:global_plots2}
\end{figure*}
%_______________________________________________________________________________________
Not only the kinesin density but also the MT density strongly affected the global patterns [Fig.\ref{fig:global_plots}].
Below a critical density $\rho_c$, the pattern was homogeneous and isotropic.
In this paper, we call it ``disordered state".
Above $\rho_c$, pattern was the LOO or LGS state depending on kinesin density.
The critical density $\rho_c$ also depended on the kinesin density, varying from 0.01 to 0.04 ${\rm\mu m ^{-2}}$. And the critical density $\rho_c$ can be quantified according to a degree of accumulation and an orientational order.

A significant feature in the LGS state is density-segregation which is characterized by the degree of accumulation calculated in the following way.
We assumed intensity of each pixel at each sliced time $I({\bf r},t)$ is linearly increased with MT density.
We calculated the static structure function $F_s(q)$ and the intermediate structure function $F_i(q,t)$ as follows:
\begin{eqnarray}
F_s(q)
&=& \langle \tilde{I}({\bf q},0)\tilde{I}({\bf q},0) \rangle _{|{\bf q}|=q}\\
F_i(q,t)
&=& \langle \tilde{I}({\bf q},t)\tilde{I}({\bf q},t) \rangle _{|{\bf q}|=q} \, %\\
\end{eqnarray}
where $\tilde{I}$ is the Fourier transform of the intensity $I$, and $\langle \cdot \rangle _{ |{\bf q}|=q }$ is ensemble average that satisfies $|{\bf q}|=q$.
Figure S12 (a) shows $F_s(q)/F_s(0)$ for each MT density. The static structure function of isolated MTs and clusters are expected to decay as a function;
\begin{eqnarray}
f(q)=\frac{A}{1+B^2q^2}
\end{eqnarray}
 (see SI \cite{SI}). The fitting coefficient $B$ represents the typical cluster size, and $A$ represents the degree of accumulation. 
We found that the parameter $A$ drastically increased at $ 0.01$ filaments par ${\rm \mu m^2}$ [Fig. \ref{fig:global_plots2}(a)], which means MT density was high enough to form cluster. We adopted 0.01 filaments par ${\rm \mu m^2}$ as the critical density for the LGS state $\rho_c^{LGS}$.

Next we investigate dynamics of clusters in this state.
Just above critical density from disorder to the LGS state, MTs form clusters crawling around.
We defined this type of cluster as ``moving cluster". As MT density increases, cluster-cluster collision in opposite directions became dominant and slowed down the speed of a merging cluster. We call this type of cluster, in particular, aggregation.
To characterize ``moving clusters'' and ``aggregation", we compared the intermediate scattering function at $0.1\  {\rm \mu m^{-1}}$ in wavelength $q$ for each MT density [Fig.\ref{fig:global_plots2}(b)]. In the moving cluster state, $F_i(q=0.1,t)$ decays fast with time $t$. Each cluster moves with a certain velocity and it makes the intermediate function decay with the following function
\begin{eqnarray}
f(t) = \frac{1}{4/Cq^2t+2/V^2q^2t^2}\ 
\end{eqnarray}
(see SI \cite{SI}). The fitting parameter $V$ corresponds to the moving velocity of the structure with the size of $1/q$.
We define the density threshold for the aggregation state as a density at which the moving velocity $V$ decreased drastically.
As shown in Fig. \ref{fig:global_plots}(b), we found that the fitting parameter $V$ drops abruptly at $0.07$ filaments par ${\rm \mu m^2}$.

At the lower kinesin density, orientation of MTs tended to be parallel for wide area. We characterized it by using the orientational order $S$, which is calculated using the following equation,
\begin{eqnarray}
S=\left \langle \left | \left \langle e^{2i \theta ({\bf r})} \right \rangle _{\bf r} \right | \right \rangle _j \  , \label{eq:S}%
\end{eqnarray}
where $\left \langle \cdot \right \rangle_{\bf r}$ is the  average over pixels in each ROI in the $j$-th image, and $\left \langle \cdot \right \rangle_{j}$ is the average over ROIs with the size of 200 ${\rm \mu m}\  \times$ 200 ${\rm \mu m}$ in the image.
We calculated orientation $\theta({\bf r})$ by OrientationJ (See SI for the details \cite{SI}). 
As shown by blue markers in Fig. \ref{fig:global_plots2} (c), $S$ significantly increased at the middle density, 0.032 and 0.050 filaments per ${\rm \mu m^2}$.
At these densities, we observed coexistence of ordered and disordered area.
A color map of the local orientational order in Fig. \ref{fig:global_plots2}(d) obviously illustrates high and low order regions spreading in bottom and top of a snapshot, respectively.
Also, the histogram of local orientational order of the same snapshot clearly shows bi-modality whose peaks are at $S= 0.4$ and 0.6 [Fig.\ref{fig:global_plots2}(c-3)]
Distributions at the both lower and higher MT densities are uni-modal [Figs. \ref{fig:global_plots2}(c-2 and 4)].
To assess the expansion of the ordered area, we measured fraction of ordered area $S>0.5$ in each snapshot at each MT density [red markers in Fig. \ref{fig:global_plots2}(a)].
The fraction was 0 below 0.036 filaments per ${\rm \mu m^2}$, and it increased abruptly at the middle density.
Based on these results, coexistence of ordered and disordered area appeared at $\rho^{CE}=0.036$ filaments per ${\rm \mu m^2}$.
In a simulation study observing orientation order of particles with uni-directional motion and nematic interaction, it is reported that such coexistence appears as nematic band \cite{Ginelli2010}.
Above 0.05 filaments per $\mu m^2$, the fraction was 1.
It means ordered area extended at least for image size 1 mm $\times$ 1 mm.
We adopted 0.05 filaments par ${\rm \mu m^2}$ as the critical density for the LOO state $\rho_c^{LOO}$. 

%-------------------------
\begin{figure*}[tbhp]
\centering
\includegraphics[width=.99\linewidth]{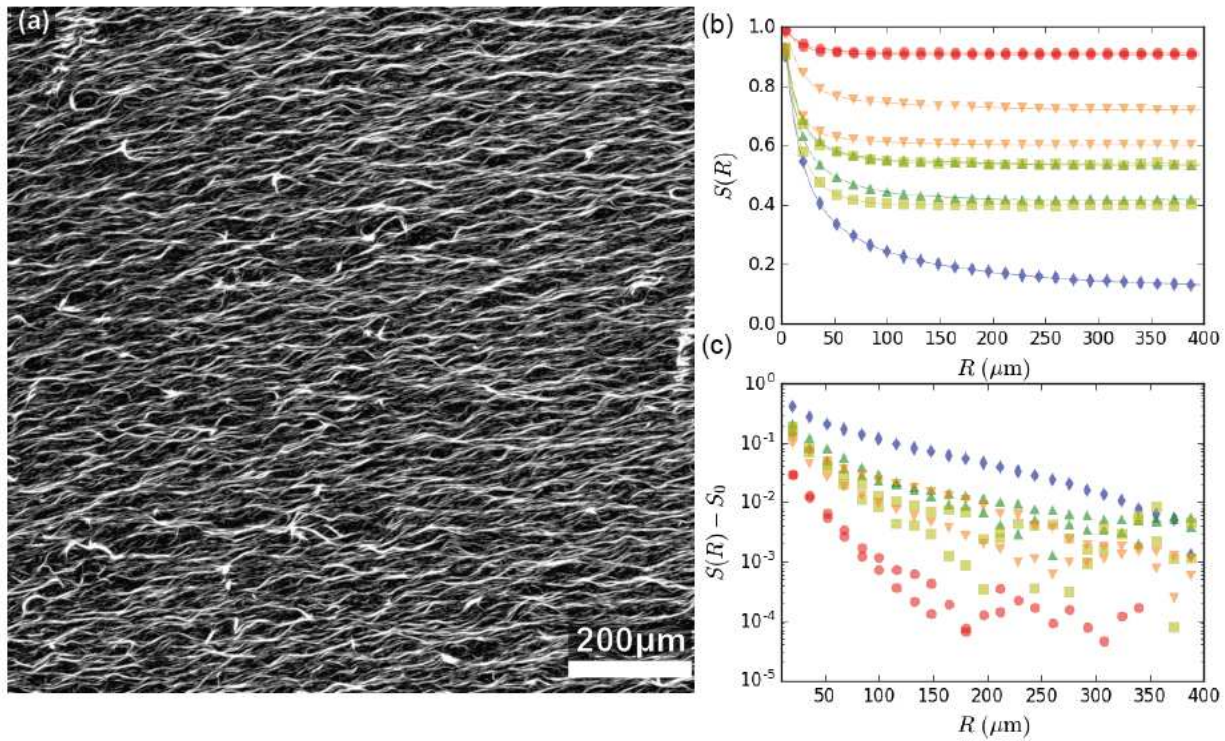}
\caption{
Long-range orientational order at the low kinesin density.
(a) A snapshot of long-range alignment of MTs at low kinesin density. Scale bar $200 \ {\rm\mu m}$ .
(b) The spacial decay of orientational order for various MT densities :
$\rho = 0.032$  (blue diamond), $0.036$  (magenta up arrow triangle), $0.040$  (yellow square), $0.050$  (orange down arrow triangle), and $0.10$  (red circle) filaments per ${\rm \mu m^2}$.
Each density has 2 data sets.
All $S(R)$ decay toward each orientation order $S_0$.
(c) $S(R)-S_0$ vs $R$ for various MT densities :
$S_0 =$ 0.13 and 0.235 ($\rho= 0.032$), $S_0 =$ 0.42 and 0.535 ($\rho= 0.036$), $S_0 =$ 0.396 and 0.54 ($\rho= 0.040$), $S_0 =$ 0.604 and 0.728 ($\rho= 0.05$), and $S_0 =$ 0.906 and 0.915 ($\rho= 0.10$).
Colors and symbols correspond to (b).
}
\label{fig:longrange}
\end{figure*}

%______________________________________________________________________

%%%%%%%%%%%%%%%%%%%%%%%%%%%%%%%
\subsubsection{Long range order of the orientational ordered pattern} \label{sec:LOOstate}
As showing in Fig. \ref{fig:longrange}(a), MTs at low-kinesin density exhibited long-range alignment across more than $1$ mm. 
To examine these patterns have long-range orientational order, we calculated orientational order for various size of area as follows;
\begin{eqnarray}
S(R) = \left\langle \left | \left \langle e^{2i \theta({\bf r}')} \right \rangle_{(|{\bf r}-{\bf r}'|<R)} \right | \right\rangle_{{\bf r}} \ ,
\end{eqnarray}
where $R$ runs from 4 to 400 $\mu$m.
When $R$ is 200 $\mu m$, $S(R)$ is the same as Eq. (\ref{eq:S}). 
At all MT densities, $S(R)$ decayed toward each asymptotic value $S_0$ [Fig. \ref{fig:longrange}(b)].
Figure \ref{fig:longrange}(c) plots $S(R) - S_0$ at each MT density.
As MT density increases, a decay of $S(R) - S_0$ became steeper. 
All of those decay curves were close to exponential.
Considering the mean MT length is about $5 \ {\rm \mu m}$ and $S(R)-S_0$ at higher density than 0.050  saturate at values larger than 0.6 before $R= 400 {\rm\mu m}$, MTs show long-range orientational order.
It is worth noting that $S(R)$ decays algebraically toward a saturated value $S_0$ instead of exponential decay, in the bacterial suspension experiment \cite{Nishiguchi2017} and numerical simulation\cite{Ginelli2010} in which true long range orders are believed to be observed for active systems with uni-directional motion and nematic alignment. In the present experiment, algebraic decay was observed only when the final pattern appeared  more disordered. 
The exponential decay toward a saturation implies the presence of a characteristic length scale. 
It might be the length scale of undulation of aligned bundles as is seen in Fig.\ref{fig:longrange}(a). The understanding the mechanism of this undulation is missing at present.

%%%%%%%%%%%%%%%%%%%%%%%%%%%%%%%%%%%%%%%%%%%%%%%%%%%%%%%%%%%%%%%%%%%%
\subsubsection{Slow rotation dynamics of the orientational ordered pattern} \label{sec:rotation}
We found that the direction of LOO rotated at around 0.05 filaments per ${\rm \mu m^2}$ [Fig. \ref{fig:rotation}(a)].
Figure \ref{fig:rotation}(b) gives the following function representing the rotation:
%___________________________
\begin{eqnarray}
{\bf n}(t) = S(t)
	\left( \begin{array}{cc}
	\cos\Theta(t)\\
	\sin\Theta(t)
	\end{array} \right)
 \ ,
\end{eqnarray}
%_________________________
where $S(t)$ is $S(R=200 \  \mu m)$ at time $t$, and $\Theta(t)$ is the direction of the LOO at time $t$.
Keeping the high orientational order $S(t)$, the global orientation of MTs $\Theta (t)$ rotates about $\pi$ radian in 6 hours.
The average velocity of MTs gliding on the lower kinesin density is about $0.2\  {\rm \mu m/s}$, so that a MT runs about 4.3 mm when the direction of LOO rotates $\pi$. 
Such rotation was observed in every area of the sample extended over more than 4 mm $\times$ 4 mm.
This chiral symmetry breaking in a global nematic ordered state could be attributed to the chirality of microtubule filaments. It is known that 14-protofilament microtubules are dominant (96\%) in MTs polymerized in the presence of GMPCPP as in our protocol \cite{Hyman1995}. 14-mer is also richer (61\%) than the normal 13-protofilament microtubules (32\%) in some standard protocols for polymerizing microtubule\cite{Ray1993}. The 14-protofilament MT has a left-handed chirality with super-twisted protofilament-lattice. If kinesin bound at the substrate moves along the protofilaments then the microtubule must rotate counterclockwise when looking in the direction of motion\cite{Ray1993}. Counterclockwise spinning MTs in their propulsion exhibit clockwise rotation on the substrate when looking from the top which corresponds to counterclockwise rotation in the inverted microscope as in the present experiment (shown in Fig.\ref{fig:rotation}). If each MT experiences clockwise torque at each point then overall nematic pattern should rotate in clockwise (counterclockwise under inverted microscope) direction synchronously over the space. This is similar to the phenomenon that swimming bacteria on the substrate exhibit clockwise rotating trajectories owing to the interaction between counterclockwise rotating flagella and the solid substrate\cite{Maeda1976,Frymier1995,Lauga2006,Hu2015}. This is also similar to Lehman rotation effect of chiral nematic liquid crystal under non-equilibrium conditions in which global pattern rotate since each molecule rotates at each position due to cross coupling effect between thermal gradient and angular speed through molecular chirality\cite{Oswald2008,Yamamoto2017,Yamamoto2018}. A weak chiral symmetry breaking in progressive MTs has been also observed in motility assay with dynein motors\cite{Sumino2012} and with kinesin\cite{Kim2018}. Although the hierarchical connection between molecular level chirality of MTs and macroscopic level is still elusive in cell chirality formation in developmental process of organisms, similar connection is well studied in macroscopic rotational motion of chiral liquid crystals due to non-equilibrium cross coupling effect\cite{Yamamoto2018}.

%-------------------------
\begin{figure}[bthp]
\centering
\includegraphics[width=.5\linewidth]{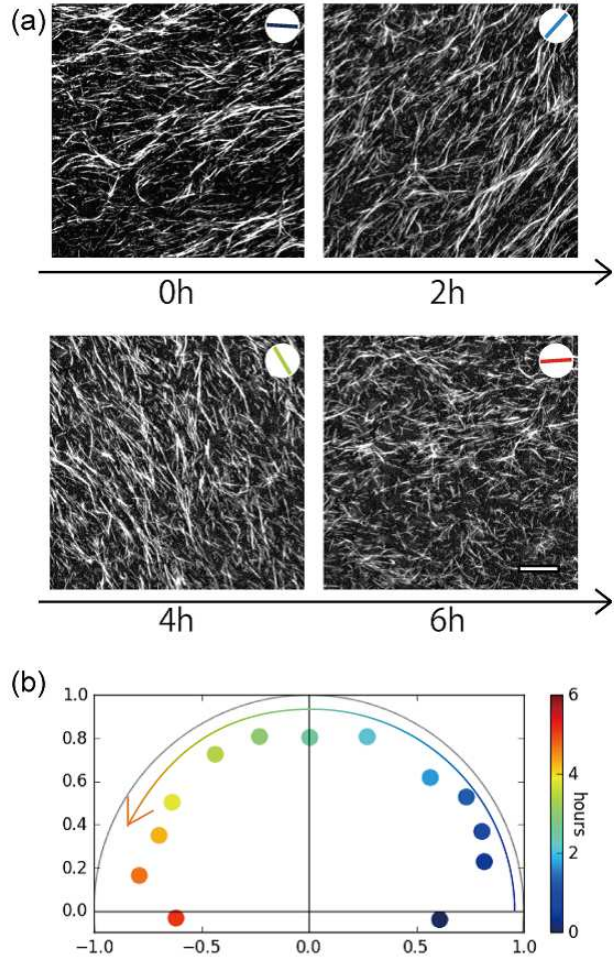}
\caption{
At about $0.05$ filaments per $\mu m^2$, the global direction of the LOO pattern rotates $\pi$ in anti-clock wise direction in $6$ hours.
(a) Snapshots of rotation of LOO pattern for every 2 hours. Bars in the white circle show the global orientation at each time. Their colors are the same as show in (b).
(b) Time evolution of the direction and magnitude of the LOO. The angle between the position of each marker and the positive x-axis shows the global direction, and the distance from the center shows the magnitude of orientational order $S$.
Color represents the time since MTs began to glide.
Scale bar $20\ {\rm \mu m}$.  
}
\label{fig:rotation}
\end{figure}
%==============================

%%%%%%%%%%%%%%%%%%%%%%%%%%%%%%%
\subsubsection{Numerical simulations}
\label{sec:numerical}

To test whether the patterns are actually determined by the strength of volume exclusion, 
we carried out numerical simulations based on the particle-based stochastic model of self-driven objects with nematic alignment and various strengths of volume exclusion. 
The numerical model based on the similar motivation was in detail investigated in Ref. \cite{Abkenar2013}, in which the rod-shaped elements are explicitly assumed and alignment between the rods is individually induced by volume exclusion. Here we propose a new numerical model as follows.
Main feature of our model is that the strengths of alignment interaction and volume exclusion can be independently controlled; namely, we here consider the case that, even when an object can cross over others easily, their directions can be aligned well with each other.
Our model also assumes anisotropic mobility so that each object hardly moves toward the direction perpendicular to its polarity, which  corresponds to the length axis of the MT in motility-assay experiments. 
This model can be mathematically formalized as given in the next paragraph, while more details of this model are found in SI \cite{SI}.

Let us assume $N$ objects in a square box with periodic boundaries in two dimensions. 
Location of the $j$-th object ${\bm x}_j =(x_j,y_j)$ ($j=1, 2, \cdots, N$) evolves over time $t$ obeying
\begin{equation} \label{eq:vdyn}
{\bm Z} ({\bm q}_j) \frac{d {\bm x}_j}{dt} = v_0 {\bm q}_j + {\bm J_j^{v}} \ ,
\end{equation}
where the vector ${\bm q}_j$ means the polarity of the $j$-th object. 
The polarity is assumed to maintain the constant magnitude, {\it i.e.} ${\bm q}_j=(\cos  \theta_j, \sin  \theta_j)$, and the direction $\theta_j$ obeys
\begin{equation} \label{eq:thetadyn}
  \frac{d \theta_j}{dt} = - {J_j^{q}}_x  \sin \theta_j + {J_j^{q}}_y \cos \theta_j + \xi_j(t) \ .
\end{equation}
Anisotropic mobility is introduced through the anisotropic friction tensor ${\bm Z} ({\bm q}_j)$ in the left hand side of Eq. (\ref{eq:vdyn}), which is defined by ${\bm Z} ({\bm q}_j)= {\bm q}_j \otimes {\bm q}_j + r_{\zeta}^{-1} ({\bm I}  - {\bm q}_j \otimes {\bm q}_j)$ 
with the ratio $r_{\zeta}=\zeta_{\parallel}/\zeta_{\perp}$ of friction coefficients in parallel $\zeta_{\parallel}$ and perpendicular directions $\zeta_{\perp}$
and the unit vector along the polarity direction ${\bm q}_j=(\cos  \theta_j, \sin  \theta_j)$.
Here, $\otimes$ mean the tensor product, and ${\bm I}$ is the identity matrix.
The first term in the right hand side of Eq. (\ref{eq:vdyn})
assumes that each object moves along its polarity ${\bm q}_j$ with a constant velocity $v_0$ in the absence of volume exclusion interactions.
The second term means mechanical volume interaction,
which is given by 
\begin{equation} \label{eq:volumeexclusion}
 {\bm J_j^{v}} = - \beta \sum_{j'} \frac{r \Delta {\bm x}_{j,j'}}{|\Delta {\bm x}_{j,j'}|^2}
\end{equation}
when $|\Delta {\bm x}_{j,j'}|<r$ with $\Delta {\bm x}_{j,j'}= {\bm x}_{j'} - {\bm x}_{j}$,
and ${\bm J_j^{v}}={\bm 0}$ otherwise.
The coefficient $\beta$ indicates the strength of volume exclusion, and $r$ is the interaction range.
The first and second terms in the right hand side of Eq. (\ref{eq:thetadyn}) 
mean nematic interaction, given by
\begin{equation} \label{eq:nematicint}
 {\bm J_j^{q}} =  2 \alpha \sum_{j'} \left( {\bm q}_{j} \cdot {\bm q}_{j'} \right) {\bm q}_{j'}
\end{equation}
when $|\Delta {\bm x}_{j,j'}|<r$ with $\Delta {\bm x}_{j,j'}= {\bm x}_{j'} - {\bm x}_{j}$, 
and ${\bm J_j^{q}}={\bm 0}$ otherwise.
The interaction range of nematic interaction is assumed here to be identical to that of mechanical volume interaction, $r$.
The coefficient $\alpha$ indicates the strength of the nematic alignment.
The last term $\xi_j(t)$ in Eq. (\ref{eq:thetadyn}) indicates the noise on the polarity, 
assumed as Gaussian white noise with $\langle \xi_j \rangle =0$ and 
\begin{equation}
\langle \xi_j (t) \xi_{j'} (t') \rangle = 2 R \delta_{j,j'}  \delta(t-t') \ ,
\label{eq:xidispersion}
\end{equation}
where $R$ is the dispersion of noise and it corresponds to the inverse correlation time of polarity direction for an isolated object. 
Time evolution of ${\bm x}_j$ and $\theta_j$ are numerically calculated 
based on Eqs. (\ref{eq:vdyn}) and (\ref{eq:thetadyn}) by the Heun's method.
Time $t$ is discretized into steps with the interval $dt = 0.004$, and
the numerical integration is carried out up to $t=2,560$. 
The parameters are set $v_0=1$, $r_{\zeta} =0.01$, $R=0.1$, $\alpha=5$ and  $r=1$.
Linear system size $L$ depends on the object density $\rho$
as $L=\sqrt{N/\rho}$ both for $x$ and $y$ directions with the given number of objects $N$.

%________________________________
\begin{figure*}[hbtp]
\centering
\includegraphics[width=.99\linewidth]{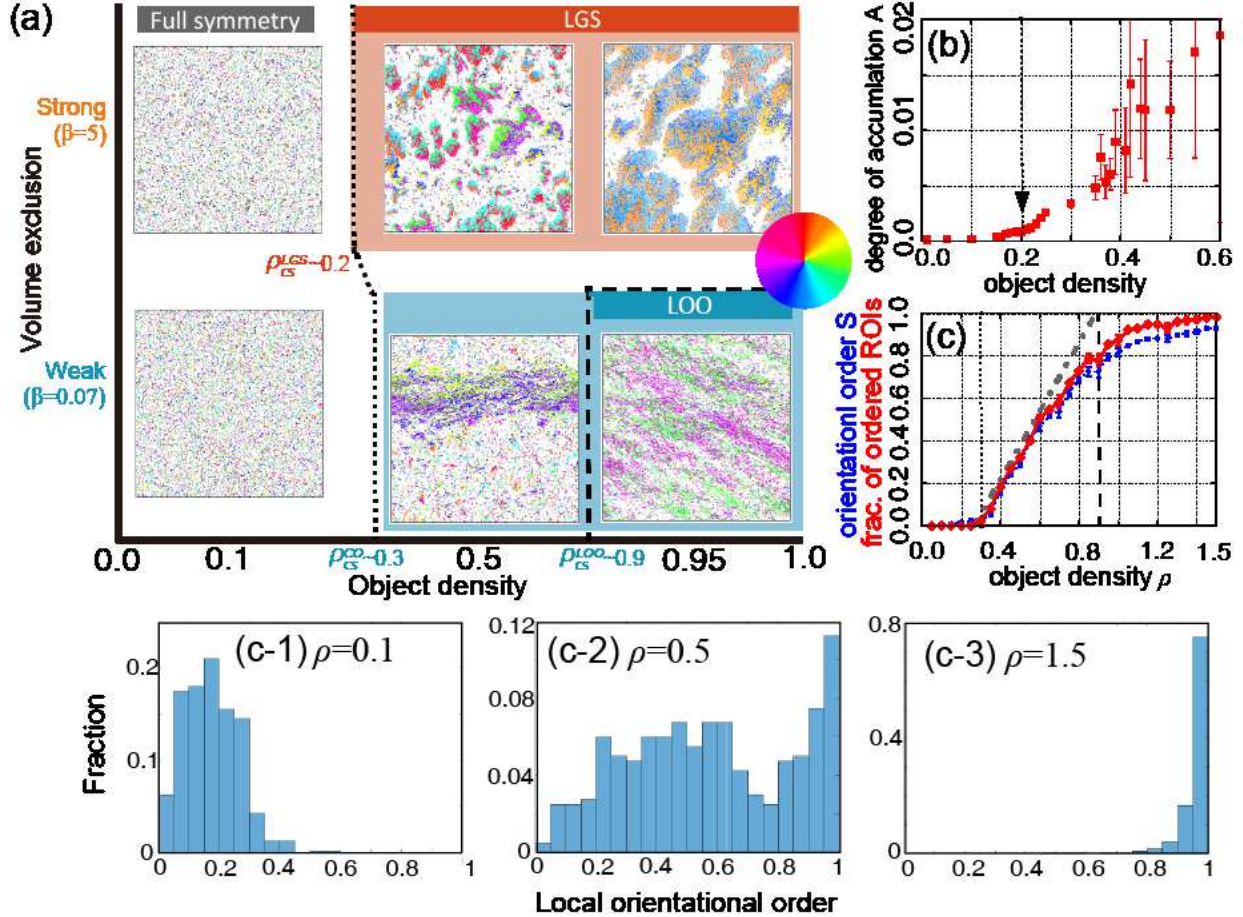}
\caption{
Numerical simulation results for collective motion of self-driven objects with volume exclusion effects, nematic alignment, and anisotropic mobility. 
(a) Phase diagram. Top and bottom rows display steady state snapshots for the cases with weak and strong volume exclusions ($\beta=0.07$ and $\beta=5.0$), respectively. The horizontal axis indicates the object density. Object density is defined in the way that interaction range is the unit of length. 
The dotted and broken lines indicate the transition line.
(b) Degree of accumulation against object density for the case with strong volume exclusion ($\beta=5.0$).
Error bars mean the error of fitting.
(c) Orientational order $S$ (blue broken line) and fractions of the ordered region (red solid line) against object density for the case with weak volume exclusion ($\beta=0.07$). 
Error bars indicate the standard error ($n=8$).
(c-1, 2 and 3) Histograms of local orientational order for $\rho=0.1$, $0.5$ and $1,5$, respectively.
The dotted and broken lines in (b) and (c) correspond to those in (a), respectively.
The gray long dashed short dashed line is the eye guide to show the increase of fraction of ordered ROIs with assuming the linear increase.
}
\label{fig:simulation}
\end{figure*}
%__________________________________

\begin{figure*}[btp]
\centering
\includegraphics[width=.95\linewidth]{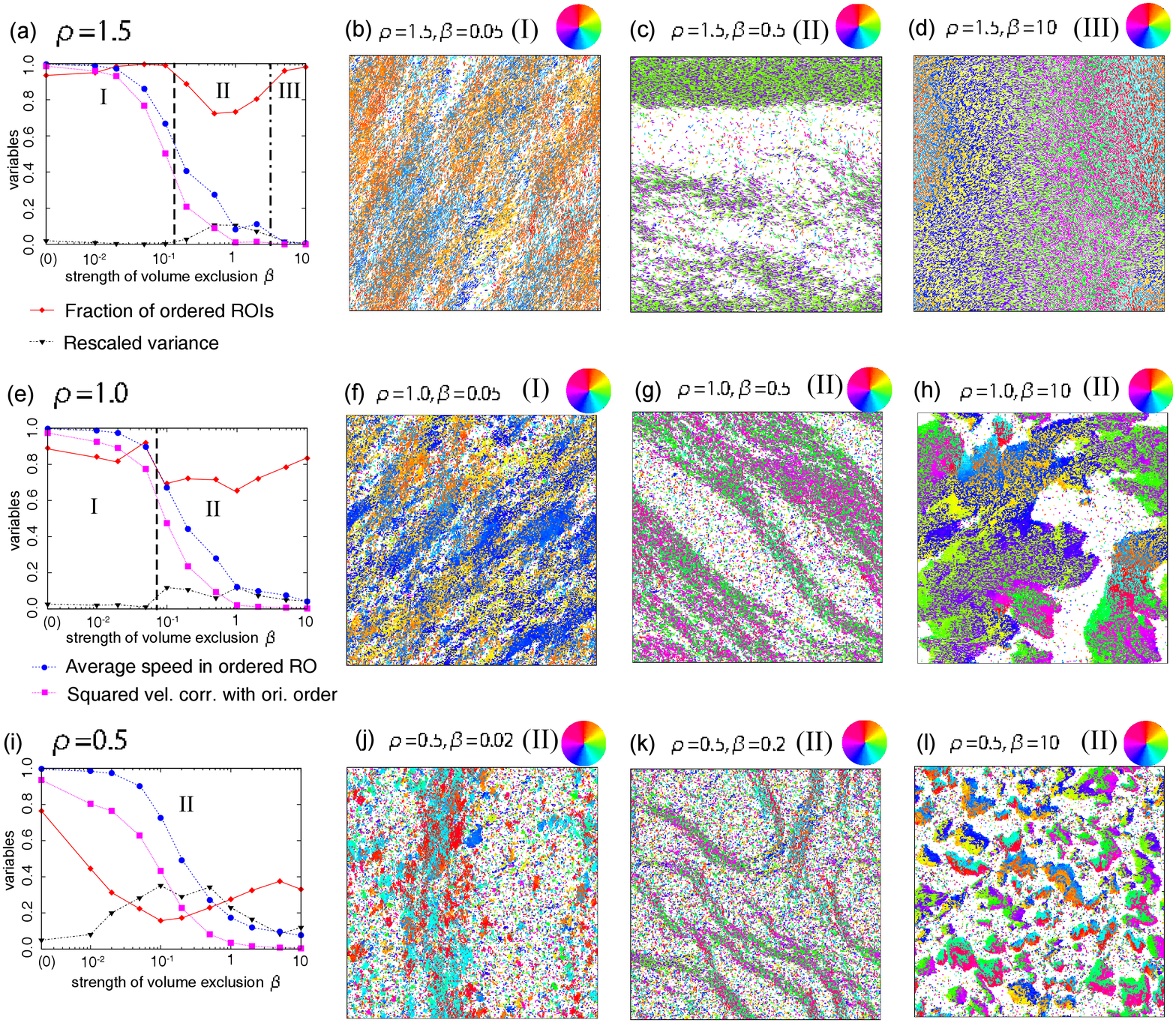}
\caption{Dependence of structure and dynamics at the steady state on volume exclusion strength in numerical results.
(a) Fraction of ordered regions, rescaled variance of the orientational order $\tilde{\rm Var}(S)$, average speed in the ordered regions, and squared velocity with correlated with orientational order $V^2_S$ against strength $\beta$ of volume exclusion for $\rho=1.5$.
Object number is fixed as $N=80,000$.
The regimes I, II and III, which correspond to the LOO, phase separated and confluent states, respectively, are identified by looking at whether $\tilde{\rm Var}(S)$ is almost zero ($\tilde{\rm Var}(S)<0.02$) or not.
(b,c,d) Snapshots at the final time step of numerical simulations for $\rho=1.5$.  
Strength of volume exclusions are set $\beta=0.02$ (b), $\beta=0.2$ (c) and $\beta=10$ (d). 
(e) The same variables in (a) for the case of $\rho=1.0$, against strength $\beta$ of volume exclusion. Object number is fixed as $N=80,000$.
In this case, for $\rho \leq 0.1$, the LOO is realized. 
 (f, g, h) Snapshots at the final time step of numerical simulations. 
Strength of volume exclusions are set $\beta=0.05$ (f), $\beta=0.5$ (g) and $\beta=10$ (h).
(i) The same variables in (a) for the case of $\rho=0.5$, against strength $\beta$ of volume exclusion. Object number is fixed as $N=80,000$.
 (j, k, l) Snapshots at the final time step of numerical simulations. 
Strength of volume exclusions are set $\beta=0.05$ (j), $\beta=0.5$ (k) and $\beta=10$ (l).
See SI for the details of analysis and more detailed $\rho-$ and $\beta-$dependence \cite{SI}.}
\label{fig:simulation2}
\end{figure*}

The results with the number of objects $N=20,000$ recapitulate our experimental observations well. 
When the strength of volume exclusion is high enough as $\beta=5$, the LGS state with motion-less aggregations is observed for high object density and disordered state is observed for low object density [Fig. \ref{fig:simulation}(a)]. 
In fact, the LGS state is quantified in Fig. \ref{fig:simulation}(b), showing the degree of accumulation $A$ at the steady state against various object density, and it shows similar tendency with experimental observation given in Fig. \ref{fig:global_plots2}(a).
The typical size of aggregation gets larger for increasing object density.
Furthermore, as shown in Figs. S2 and S3, at $\rho > 1$ for $\beta \geq 5$ the almost entire space is filled with objects, where topological defects and their slow pair annihilation dynamics are observed (See SI \cite{SI}; see Fig. \ref{fig:simulation2}(d) also).
In contrast, when objects had low strength of volume exclusion as $\beta=0.07$, the LOO, nematic band and disordered state were observed depending on the object density.
The LOO state is quantified through global orientational order $S$ and fraction ordered ROIs and histograms of local orientational order at the steady state against various object density in Fig. \ref{fig:simulation2}(c), which agrees well with the experimental result in Fig. \ref{fig:global_plots}(c). 
Note that it is known that the orientational order becomes zero for larger system size since the bending mode of the stream is unstable in the phase coexistence regime or the nematic band \cite{Ginelli2010}.

To examine crossing over from the LOO to the LGS when the volume exclusion effect increases,
we investigated the structure and dynamics at the steady state for various volume exclusion strength $\beta$ in Fig. \ref{fig:simulation2} 
for $\rho=1.5$, $\rho=1.0$ and $\rho=0.5$ (with $N=40,000$, $N=80,000$ and $N=80,000$, resp.).
To quantify the structure, we evaluated fraction of the region where orientational order is high as $S>0.75$ and variance of the order parameter, as shown by the red squares in Fig. \ref{fig:simulation2}(a) for firstly $\rho=1.5$.
The fraction of ordered region is almost $1$ for $\beta \leq 0.1$, whereas at $\beta \geq 0.2$, it becomes smaller and deviates from $1$ [Fig. \ref{fig:simulation2}(a); red diamonds].
Further increase of $\beta$ for $\beta \geq 1$ leads to the increase of the fraction, and for $\beta \geq 5$ we find that it the ordered fraction gets roughly $1$ again.
The typical snapshots at each value of $\beta$ are illustrated in  Fig. \ref{fig:simulation2}(b)-(d): 
For $\beta=0.05$, the LOO state was observed [Fig. \ref{fig:simulation2}(b)], which was the same as a pattern when $\beta=0.07$ and object density was 0.95 in Fig. \ref{fig:simulation}(a).
In contrast, for $\beta=0.5$, the density phase separation occurs [Fig. \ref{fig:simulation2}(c)]. 
Furthermore, for $\beta = 10$, the entire system gets filled with objects [Fig. \ref{fig:simulation2}(d)] because the objects become unable to overlap with each other [Fig. \ref{fig:simulation2}(a); $\beta \geq 5$]. We refer to this state as confluent state in this article.
The same conclusion is obtained by evaluating the rescaled variance of orientational order $\tilde{\rm Var}(S) = (\Delta S^{\rm ROI})^2 / \langle S^{\rm ROI} \rangle^2_{{\rm ROI},t}$ with the orientational order $S^{\rm ROI}$ defined in each ROI and its variance $(\Delta S^{\rm ROI})^2$ over various ROIs and time [Fig. \ref{fig:simulation2}(a); black inverse triangles]. We identified the regimes I, II and III in Fig. \ref{fig:simulation2}(a) by looking at whether $\tilde{\rm Var}(S)$ is almost zero or not. Each regime corresponds to the LOO, phase separated and confluent states, respectively.
Next we investigated the case for $\rho=1$ as shown in Figs. \ref{fig:simulation2}(e)-(h).
We found that, also in this case, the homogeneous order is violated when $\beta$ is increased larger than $0.1$ [see red circles in Fig. \ref{fig:simulation2}(e)]. 
However, the confluent state is not observed even at $\beta =10$ [Fig. \ref{fig:simulation2}(h)].
The same analyses were performed also for $\rho=0.5$, as shown in Figs. \ref{fig:simulation2}(i)-(l).
The fraction of ordered region decreases for increasing $\beta$ at $\beta < 0.2$, whereas at $\beta \geq 0.5$, it takes around $0.3$.  Again note that the bending mode of this band is known to become unstable with much larger system size \cite{Ginelli2010}. 
When one increases $\beta$ up to $0.2$, the band become much narrower and the bending mode becomes unstable within smaller length scale [Fig. \ref{fig:simulation2}(k)]. For larger $\beta$ like $\beta=10$, the band itself gets disassembled [Fig. \ref{fig:simulation2}(l)].
We also investigated the dynamics by firstly evaluating the average speed of objects in ordered regions.
In what follows, we focus on Figs. \ref{fig:simulation2}(i)-(l).
As shown by blue circles in Fig. \ref{fig:simulation2}(i), the average speed gradually decreases as the volume exclusion strength $\beta$ is increased.
The average speed stays finite even for $\beta=10$, where the snapshot already shows the aggregation state.
This can be because that objects in an aggregation can slowly move into the direction perpendicular to that of orientational order, since the anisotropy of friction $r_{\zeta}$ is finite here.
To see only the object motion along the direction of orientation order of each aggregation, 
we also plotted the squared velocity correlated with orientational order 
\begin{equation}
V^2_S = \frac{\langle  (1/N) \sum_{k {\rm : ROI}} \sum_{ij} S_{ij}^{\rm ROI} v_i^k v_j^k \rangle_{\rm ROI}}{\langle n^{\rm ROI} S^{\rm ROI} \rangle_{\rm ROI}}
\end{equation}
[The magenta squares in Fig. \ref{fig:simulation2}(i)]
with the object number fraction $n^{\rm ROI} \equiv (1/N) \sum_{k {\rm : ROI}} 1$
and the tensor orientational order $S_{ij}^{\rm ROI} \equiv \sum_{k {\rm : ROI}} 2 [q_i q_j - (1/2) \delta_{ij}]$ ($i=x,y$, $j=x,y$) in each ROI,
and the velocity vector $v_i^k$ ($i=x,y$) of each object ($k$-th object).
The summation $\sum_{k {\rm : ROI}}$ ran over objects in the ROI, and the average $\langle \cdot \rangle_{\rm ROI}$ was taken over the various ROI locations.
Also, $S^{\rm ROI}$ was the orientation order defined in the ROI, which was identical with the positive eigenvalue of the tensor order parameter $S_{ij}^{\rm ROI}$.
This variable $V^2_S$ is also showing the gradual decrease as $\beta$ is increased, and at $\beta \gg 0.1$, asymptotically goes to zero for increasing $\beta$.
The similar shapes of curves are obtained also in Figs. \ref{fig:simulation2}(a) and (e).

%%%%%%%%%%%%%%%%%%%%%%%%%%%%%%%
\subsection{Statistics of binary collisions} \label{sec:binarycollision}
In the previous subsections, we found various collective patterns of MTs 
like the LOO and LGS states 
depending on the strength of volume exclusion.
To understand the mechanism how the volume exclusion affects the emerging patterns, we examined the statistics of binary collisions between MTs.
We decreased the MT density to $0.4\times 10^{-3}$ filaments per ${\rm \mu m^2}$ to focus on only pair interactions between isolated MTs experimentally.
When two filaments encountered each other, they exhibited three types of behaviors: crossing-over, alignment or anti-alignment, as already shown in Figs. \ref{fig:snaps}(a)-(c).

To quantify consequence of binary collision, we measured incoming and outgoing angles defined in Sec. \ref{sec:strength} and duration time $\tau$ for the lower and higher kinesin density.
The duration time is defined as a time between two MTs touch to detach.
In order to characterize degree of alignment just after collisions independently of other factor like fluctuation of moving direction, we especially defined perfect alignment and anti-alignment assigned $\theta_{out}=0$ and $\pi$ respectively as the followings:
(1) events in which the one of colliding MTs becomes parallel and anti-parallel to the other, respectively, (2) events whose $\tau$ is longer than the time by which a MT glides 4 ${\rm\mu m}$ and the incoming angle $\theta_{in}$ is smaller and larger than $\pi/2$, respectively.
As shown by the red lines in Fig. \ref{fig:binary}(a), which indicate averages of the outgoing angles for every $\pi/9$ incoming angles, binary collision gives rise to nematic alignment at the lower kinesin density.
In contrast, at the higher kinesin density, binary collisions with obtuse incoming angles exhibited effective polar alignment rather than nematic alignment [Fig. \ref{fig:binary}(b)].

Based on this observation, we slightly modified the numerical model given in the previous subsection to recapitulate effective polar alignment tendency given in Fig. \ref{fig:binary}(b).
Simulation results with this model are shown in Fig. S8, which actually shows the moving clusters.

In comparing situations at high and low kinesin densities,
a clear difference appears in distribution of the duration time of collision $\tau$ defined above.
At the higher kinesin density, the acute incoming angle tends to show longer duration time $\tau$ than that of the obtuse incoming angles while $\tau$ at the lower kinesin density did not exhibit a significant difference between acute and obtuse incoming angles [Fig. \ref{fig:binary}(c)].
To qualify this difference, we define the following function:
\begin{eqnarray}
\label{eq:t}
	T=\frac{\langle\tau(\{\theta_{in}\}:\theta_{in}\leq \pi/6)\rangle-\langle\tau(\{\theta_{in}\}:\theta_{in}\geq 5\pi/6)\rangle}{\langle\tau(\{\theta_{in}\})\rangle}
\end{eqnarray}
where $\tau(\{\theta_{in}\})$ is mean duration time of collision for a set of ${\theta_{in} }$.
As obviously indicated by $T$ in Fig. \ref{fig:binary}(d), only the $\tau$ at the higher kinesin density had a difference.

Based on these results, we assumed that asymmetry of duration time $\tau$ produced the coherent motions of MTs in a single cluster.
If a MT collides at the pair of MTs, three MTs are aligned to be  parallel and then the middle one, which is sandwiched by the others, cannot escape from the group [Fig. \ref{fig:binary}(e)].
Thus, a condition of cluster formation can be estimated by the duration time $\tau$ and collision frequency $z$.
When  $z$ is larger than $1/\tau$, clusters are formed.
We assume the MT density $\rho$ is uniform at the initial state, and $z$ is represented by the following equation:
\begin{eqnarray} \label{eq:colfre}
z = 2 \ell v_m \rho/\pi \ ,
\end{eqnarray}
where $\ell$ refers to a typical length of a MT, and $v_m$ is a typical velocity of a MT.
One can obtain the condition for the cluster formation by replacing $z$ by the inverse duration time $1/\tau$ in Eq. (\ref{eq:colfre}).
We obtained the lowest density to form cluster, $\rho_c^{\rm est}$, to be $\sim 0.02$ filaments per ${\rm}$ using the real values of parameters, $\tau \sim$ 300 s, $\ell \sim 5\  {\rm \mu m}$, and $v_m \sim 0.05 \  {\rm \mu m /s}$.
This value of $\rho_c^{\rm est}$ is comparable with the experimental result, $\rho_c^{LGS}= 0.01$ filaments per ${\rm \mu m^2}$.
It appears that a long accompaniment after collide in acute angle causes polar motion of a cluster. 
Note that the kinesin density also affects velocity of MTs.
Speed at low kinesin density is 0.2 ${\rm \mu m/s}$ and that at high kinesin density is 0.05 ${\rm \mu m/s}$.
Although MTs run slower as the kinesin density increases, it dose not change the consequence of this discussion.

\begin{figure*}[thbp]
\centering
\includegraphics[width=.95\linewidth]{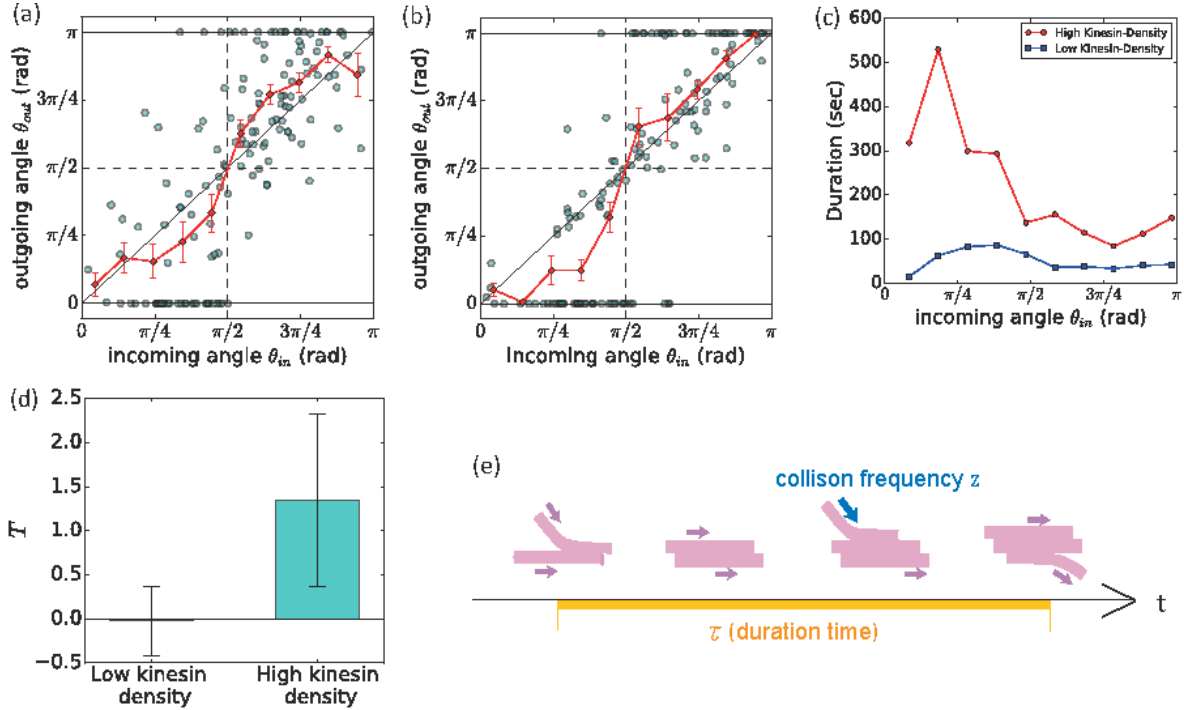}
\caption{(a,b)Effects of binary collision of MTs.  Angular relation between two MTs before and after binary collision when kinesin density was lower (a, number of collision events $n=147$) and when it was higher (b, $n=135$).   Each dot represents each collision event. A red line represents mean and mean error of alignment angles for every $\pi/9$ degree of incoming angles. (c) Duration time of collision $\tau$ from touch to detach for every $\pi/9$ degree of incoming angles. Red circle and blue square markers represent durations at the higher and lower kinesin densities, respectively. (d) Effective difference between duration of acute ($\leq\pi/6$) and obtuse ($\geq5\pi/6$) incoming angles. (e) Schematic image of cluster formation.}
\label{fig:binary}
\end{figure*}
%%%%%%%%%%%%%%%%%%%%%%%%%%%%%%%%%

\subsection{Loop formation}\label{sec:loop}
As we discussed in Sec. \ref{sec:binarycollision}, clear difference of duration time between alignment and anti-alignment led to unidirectional motion of each cluster at high kinesin density.
After forming clusters, cluster-cluster collisions became dominant as shown in the areas surrounded by blue lines in Fig. \ref{fig:defects} and SI movie 2 \cite{SI}.
Also, some moving clusters changed its direction of motion and then formed loops and arcs to which all MTs belonging rotate in the same direction [Figs. \ref{fig:defects}(a)-(c)].
The average inner diameter was 4.2 ${\rm \mu m}$. 
These loops remind us of cytoskeletal arcs in the gliding dynamics of single filaments \cite{Kawamura2008,Ziebert2015,Liu2011b,Kabir2012}, bundles \cite{HenryHess2005,Schaller2011,Tamura2011,Inoue2013,Lam2014a}, and in collective motion \cite{Sumino2012}.
Those can be characterized by loop size and rotation direction of filaments in a loop.
Typical diameters of arcs of single filaments, bundles and collective motions were about 2, 5 and 400 ${\rm \mu m}$, respectively.
Rotation directions of single filament and bundle loops were unidirectional, while that of collective motions was bidirectional.
Loops in our experiments are similar to the bundle loop in size and rotation direction. 
Here, we show the fact that the model proposed above can recapitulate this loop structure by adding a small change.
We added the interaction term expressing the contact following into Eq. (\ref{eq:nematicint}), which is now rewritten by
\begin{eqnarray}
 {\bm J^{q}}_j &=& 2 \alpha \sum_{j'} \left( {\bm q}_{j} \cdot {\bm q}_{j'} \right) {\bm q}_{j'}
 + \alpha_{\rm CF} \sum_{j'}  
 \frac{1}{4} \left( 1+{\bm q}_{j} \cdot \frac{\Delta {\bm x}_{j,j'}}{|\Delta {\bm x}_{j,j'}|} \right) \nonumber  \\ 
 &\times& \left( 1+{\bm q}_{j'} \cdot \frac{\Delta {\bm x}_{j,j'}}{|\Delta {\bm x}_{j,j'}|} \right) \frac{\Delta {\bm x}_{j,j'}}{|\Delta {\bm x}_{j,j'}|} \ ,
\label{eq:nematicintwithCF}
\end{eqnarray}
reflecting the following observations. 
When two microtubules collided with each other, their direction after the collision did not always get the average direction of in-coming directions.
Rather, seemingly the MT bumping into the backside of another MT had the tendency to follow it.
The result is demonstrated in Fig. \ref{fig:defects}(d)
for $\alpha_{\rm CF}=1.0$, $\alpha=1.0$, $\beta=0.05$  and $\rho=0.5$ (the other parameters are the same as above),
which actually shows the loop structure.
However, this loop structure seems shrinking and unstable for longer time. 
The similar loop structure is observed transiently in the numerical simulation of the self-propelled system with the visual corn \cite{Barberis2016,Peruani2017}.

\begin{figure*}[thbp]
\centering
\includegraphics[width=.9\linewidth]{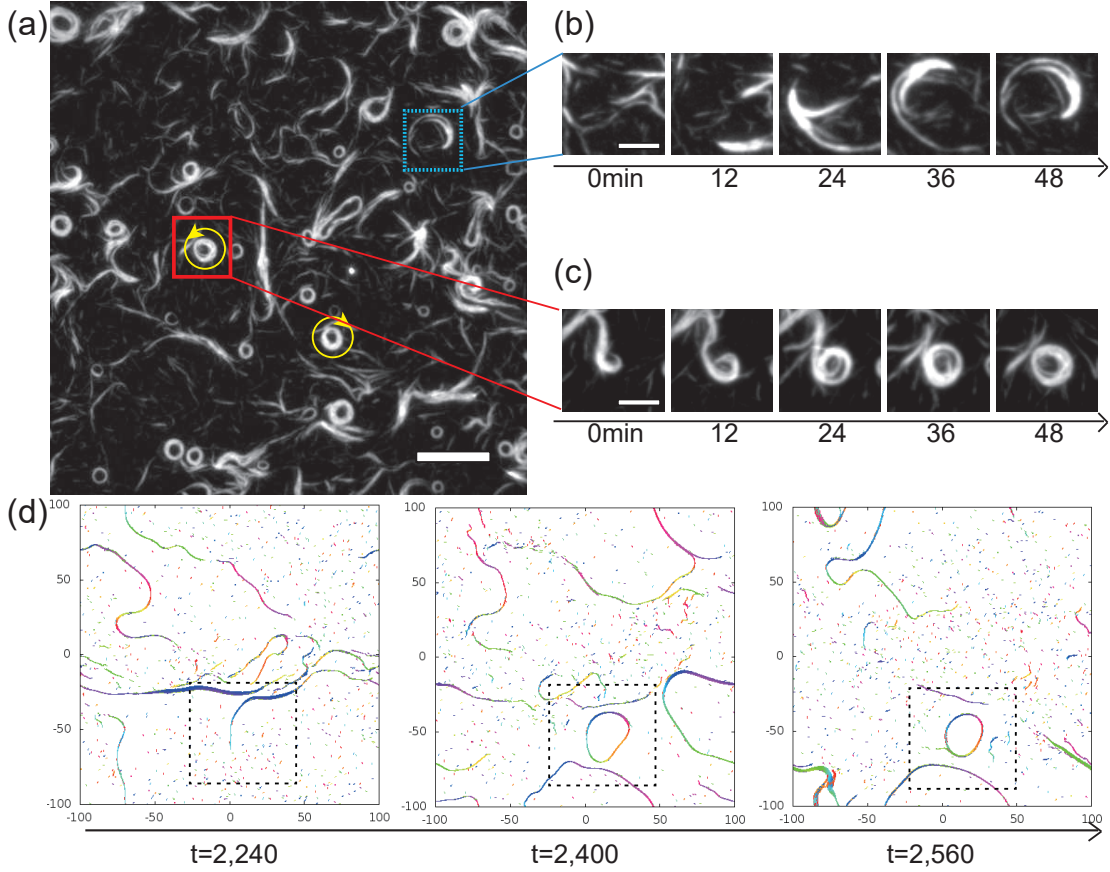}
\caption{
(a-c) Snapshots of moving clusters and loops at the higher kinesin density $6.1 \times 10^3$ per ${\rm \mu m^2}$.
	MT density is the higher than the density threshold for aggregation pattern, 0.22 filaments per ${\rm \mu m^2}$.
	This induces frequent cluster-cluster collisions [in a blue square in (a)] and results in formations of loops whose rotation directions are random as indicated by the yellow circles with arrows in (a). Scale bar is $20{\rm \mu m}$. Time evolution of loop at the area surrounded by blue (b) and red (c) square in the snapshot (a). 	A tip of a cluster were bended to collide with itself and caught in 4 ${\rm \mu m}$ inner diameter of a loop.  The snapshot (a) correspond to 48 min after this loop formation. Scale bar is $10\ {\rm \mu m}$.
(d) Snapshots from the numerical simulation showing the loop formation. The numerical model with contact following was used. See the main text for details.
}
\label{fig:defects}
\end{figure*}

\section{Discussion}
\label{sec:discussion}
This study reasoned the effect of volume exclusion on collective motion whose symmetry is polar motion and nematic alignment, and found that excessively strong volume exclusion works to destroy orientation order and form clusters. 
Our findings explain previous works \cite{Hussain2013a,Sumino2012,Inoue2015,Nishiguchi2017,Peruani2012,Zhang2010} according to the volume exclusion effect.
The probability of crossing in the motility assay with actins and myosins by S. Hussain et al \cite{Hussain2013a} is about $50\%$, and that with MTs and dyneins by Y. Sumino et al. \cite{Sumino2012} is about $20\%$, and that with microtubules and kinesins adding methylcellulose by Inoue et al. \cite{Inoue2015} is about $50\%$.
All these systems give high orientational order. 
Also, elongated {\it E. coli} confined in a quasi-two-dimensional chamber, where {\it E. coli} can cross over with each other, also showed the true long-range orientational order \cite{Nishiguchi2017}.
These results can be interpreted as the consequence that high crossing-over probability and alignment interaction can form orientational order.
In contrast, the systems in which crossing-over is prohibited, like Myxobacteria by F. Peruani et al. \cite{Peruani2012} and {\it B.subtilis}  by H. P. Zhang et al. \cite{Zhang2010}, formed the moving cluster pattern and did not exhibit any orientational order.
Our results are consistent with these results, {\it i.e.} orientational order emerges at low kinesin density where MTs are able to cross over, and LGS emerge at high density where they cannot cross over.

It may seem strange that lower orientation order is made by stronger excluded volume effects, by which one may expect also stronger alignment effect. 
According to statistics of binary collisions, nematic interaction dose not disappear even when volume exclusion is weaker. Based on this, our numerical model can implement the strength of alignment independently from the volume exclusion strength.

Furthermore, previous insight and our results in Fig. \ref{fig:simulation2}(d) suggest that highly packed self-propelled rods show high orientational order \cite{Samuel2012,Abkenar2013}.
When area fractions of particles are close to 1, every particles are surrounded by others and aligned to the same direction.
The mechanism to form these ordered patterns are common to highly packed rods including those reverting their motile direction occasionally, for example, fluidized granular rods, myxobacteria, and neural stem cells \cite{Narayan2007a,Wu2009,Kawaguchi2017}. 
Theoretically, this high orientation order is considered as a quasi long range order\cite{Chate2006} for so called ``active nematics" with bi-directional motion with nematic interaction.
Difference between uni- and bi-directional motion would appear in the regime of middle area fraction.
When there is enough room to move around, unidirectional motile particles gather and form asters and vortices \cite{Nedelec1997,Kruse2005a,Backouche2006,Torisawa2016a}.
On the other hand, bidirectional motile particles can escape from such aster and vortex structures by reverting its direction of motion.
It would be assumed that reversal time of direction of motion would be a parameter determining whether these structures are stable or not.

Critical densities obtained in experiments can be compared with those in numerical simulations.
The length scale of experimental results is normalized by a half of the typical MT length $\ell/2 \sim 2.5 \  {\rm \mu m}$.
As a result, the normalized critical densities from disorder to ``liquid-gas-like phase separation (LGS)" state $\rho_c^{\rm LGS}$ and to the coexistence of ordered and disordered phases $\rho_c^{\rm CO}$ become $0.06$ and $0.23$, respectively.
These are of the same order of magnitude as those obtained in numerical simulations, $\rho_{cs}^{\rm LGS}=0.2$ and $\rho_{cs}^{\rm CO}=0.3$. The magnitude relation is also the same between them as $\rho_c^{\rm LGS}<\rho_c^{\rm CO}$ and $\rho_{cs}^{\rm LGS}<\rho_{cs}^{\rm CO}$.

In experiment, various striking patterns appeared.
Loops and moving cluster could be due to the contact following which appear to act when the volume exclusion is high.
Actually, as shown in Sec.\ref{sec:loop}, contact following can yield loops and moving cluster and keep them for a while.
In addition, loops are the typical structure of unidirectional motile rods with strong volume exclusion at middle density. 
Comparing their size and rotation direction, loops in our experiment are similar to the bundle loops in previous studies \cite{HenryHess2005,Schaller2011,Kakugo2011,Tamura2011,Inoue2013,Lam2014a}.
However, origins of alignment interaction are different.
Previous studies used cross-linkers while we confined MTs to  increased the kinesin density without cross-linkers. 
Although interaction details are different, we believe almost the same mechanism to form loops are working because alignment behaviors of colliding filaments are common.

\section{Conclusions}
\label{sec:conclusions}
In summary, we focused on the effect of volume exclusion on global patterns of collective motions with motility assay using microtubule (MT) and kinesin in order to give a unified view for patterns reported in previous studies.
We varied the strength of volume exclusion between MTs, i.e. the probability of MT overlapping in collision, by changing the kinesin density. MTs at low kinesin density could cross over. Meanwhile, those at high kinesin density did not cross over, rather often show perfect alignment more frequently.
Surprisingly, however, we found that the orientational order emerged when volume exclusion is weak, whereas the cluster patterns without the orientational order emerged when volume exclusion was strong.
We also found the MT density-dependency of patterns.
Firstly, near the critical density, static appearances of MTs change from disordered to patterned states, which includes the nematic band or the coexistence of ordered and disordered phases, ``long-range orientational order (LOO)" and  ``liquid-gas-like phase separation (LGS)" states.
The critical density was found to depend on the kinesin density.
Secondly, dynamical appearance of MTs in the LGS state changed from ``moving clusters'' to ``aggregation" when we increased MT density.
Numerical simulation confirmed that whether the system is in disordered or patterned states depends on the object density.
Moreover it could recapture characteristics of patterns in experiment, {\it i.e.} the coexistence of ordered and disordered phases, the LOO and LGS states, under proper strength of volume exclusion.

To our best knowledge, this study reports the first experimental results showing the effect of volume exclusion directly.
It shows that controlling the strength of volume exclusion can produce both long-range alignment and clusters.
Although our study focused on experimental systems where elements move unidirectional and interact nematically, this approach is expected to be applied to other sets of symmetry of motion and interaction in future works.

\section{Acknowledgment}
We gratefully acknowledge the experimental work of past and present members of Higuchi laboratory.
We are grateful to Hideo Higuchi, Motoshi Kaya, Hugues Chate, Ken H Nagai, Daiki Nishiguchi
and Natsuhiko Yoshinaga
for helpful discussions and for their kind interest in this work. We would like to thank Zvonimir Dogic for practical advice for experiment.
This work is ostensibly supported by JSPS KAKENHI Grant Number 25103004, 16H02212, JP16J06301 and JP16K17777.

%\bibliographystyle{abbrv}
%\bibliography{lib}

\begin{thebibliography}{70}

\bibitem{SI}
See Supplemental Information at http://link.aps.org/supplemental/*** for
  details on the theory behind the analysis using scattering functions, the
  models and results of numerical simulations, and the image processing and
  analysis of experimental data. Supplemental Material also includes movies
  from experimental results.

\bibitem{Abkenar2013}
M.~Abkenar, K.~Marx, T.~Auth, and G.~Gompper.
\newblock {Collective behavior of penetrable self-propelled rods in two
  dimensions}.
\newblock {\em Physical Review E}, 88(6):062314, dec 2013.

\bibitem{Backouche2006}
F.~Backouche, L.~Haviv, D.~Groswasser, and A.~Bernheim-Groswasser.
\newblock {Active gels: dynamics of patterning and self-organization}.
\newblock {\em Physical Biology}, 3(4):264--273, dec 2006.

\bibitem{Ballerini2008a}
M.~Ballerini, N.~Cabibbo, R.~Candelier, A.~Cavagna, E.~Cisbani, I.~Giardina,
  V.~Lecomte, A.~Orlandi, G.~Parisi, A.~Procaccini, M.~Viale, and
  V.~Zdravkovic.
\newblock {Interaction ruling animal collective behavior depends on topological
  rather than metric distance: evidence from a field study.}
\newblock {\em Proceedings of the National Academy of Sciences of the United
  States of America}, 105(4):1232--7, jan 2008.

\bibitem{Barberis2016}
L.~Barberis and F.~Peruani.
\newblock {Large-Scale Patterns in a Minimal Cognitive Flocking Model:
  Incidental Leaders, Nematic Patterns, and Aggregates}.
\newblock {\em Physical Review Letters}, 117(24):248001, dec 2016.

\bibitem{Bricard2013}
A.~Bricard, J.-B. Caussin, N.~Desreumaux, O.~Dauchot, and D.~Bartolo.
\newblock {Emergence of macroscopic directed motion in populations of motile
  colloids}.
\newblock {\em Nature}, 503(7474):95--98, nov 2013.

\bibitem{Butt2010}
T.~Butt, T.~Mufti, A.~Humayun, P.~B. Rosenthal, S.~Khan, S.~Khan, and J.~E.
  Molloy.
\newblock {Myosin motors drive long range alignment of actin filaments}.
\newblock {\em Journal of Biological Chemistry}, 285(7):4964--4974, feb 2010.

\bibitem{Castoldi2003}
M.~Castoldi and A.~V. Popov.
\newblock {Purification of brain tubulin through two cycles of
  polymerization–depolymerization in a high-molarity buffer}.
\newblock {\em Protein Expression and Purification}, 32(1):83--88, nov 2003.

\bibitem{Chate2006}
H.~Chat{\'{e}}, F.~Ginelli, and R.~Montagne.
\newblock {Simple Model for Active Nematics: Quasi-Long-Range Order and Giant
  Fluctuations}.
\newblock {\em Physical Review Letters}, 96(18):180602, may 2006.

\bibitem{DeCamp2015}
S.~J. DeCamp, G.~S. Redner, A.~Baskaran, M.~F. Hagan, and Z.~Dogic.
\newblock {Orientational order of motile defects in active nematics}.
\newblock {\em Nature Materials}, 14(11):1110--1115, 2015.

\bibitem{Frymier1995}
P.~D. Frymier, R.~M. Ford, H.~C. Berg, and P.~T. Cummings.
\newblock {Three-dimensional tracking of motile bacteria near a solid planar
  surface.}
\newblock {\em Proceedings of the National Academy of Sciences of the United
  States of America}, 92(13):6195--9, jun 1995.

\bibitem{Furuta2017}
A.~Furuta, M.~Amino, M.~Yoshio, K.~Oiwa, H.~Kojima, and K.~Furuta.
\newblock {Creating biomolecular motors based on dynein and actin-binding
  proteins}.
\newblock {\em Nature Nanotechnology}, 12(3):233--237, nov 2017.

\bibitem{Furuta2013}
K.~Furuta, A.~Furuta, Y.~Y. Toyoshima, M.~Amino, K.~Oiwa, and H.~Kojima.
\newblock {Measuring collective transport by defined numbers of processive and
  nonprocessive kinesin motors.}
\newblock {\em Proceedings of the National Academy of Sciences of the United
  States of America}, 110(2):501--6, 2013.

\bibitem{Gerum2013}
R.~C. Gerum, B.~Fabry, C.~Metzner, M.~Beaulieu, A.~Ancel, and D.~P. Zitterbart.
\newblock {The origin of traveling waves in an emperor penguin huddle}.
\newblock {\em New Journal of Physics}, 15(12):125022, dec 2013.

\bibitem{Ginelli2010}
F.~Ginelli, F.~Peruani, M.~B{\"{a}}r, and H.~Chat{\'{e}}.
\newblock {Large-Scale Collective Properties of Self-Propelled Rods}.
\newblock {\em Physical Review Letters}, 104(18):184502, may 2010.

\bibitem{Gregoire2004}
G.~Gr{\'{e}}goire and H.~Chat{\'{e}}.
\newblock {Onset of Collective and Cohesive Motion}.
\newblock {\em Physical Review Letters}, 92(2):025702, jan 2004.

\bibitem{Helbing2000}
D.~Helbing, I.~Farkas, and T.~Vicsek.
\newblock {Simulating dynamical features of escape panic}.
\newblock {\em Nature}, 407(6803):487--490, sep 2000.

\bibitem{HenryHess2005}
H.~Hess, C.~John, B.~Christian, R.~Doot, S.~Luna, E.~Karl-Heinz, and V.~Vogel.
\newblock {Molecular Self-Assembly of ``Nanowires'' and ``Nanospools''
  Using Active Transport}.
\newblock {\em Nano Letters}, 5(4):629--633, 2005.

\bibitem{Hu2015}
J.~Hu, A.~Wysocki, R.~G. Winkler, and G.~Gompper.
\newblock {Physical Sensing of Surface Properties by Microswimmers - Directing Bacterial Motion via Wall Slip}.
\newblock {\em Scientific Reports}, 5(1):9586, aug 2015.

\bibitem{Hussain2013a}
S.~Hussain, J.~E. Molloy, and S.~M. Khan.
\newblock {Spatiotemporal dynamics of actomyosin networks}.
\newblock {\em Biophysical Journal}, 105(6):1456--1465, 2013.

\bibitem{Hyman1991}
A.~Hyman, D.~Drechsel, D.~Kellogg, S.~Salser, K.~Sawin, P.~Steffen,
  L.~Wordeman, and T.~Mitchison.
\newblock {Preparation of modified tubulins}.
\newblock {\em Methods in Enzymology}, 196(C):478--485, 1991.

\bibitem{Hyman1995}
A.~A. Hyman, D.~Chr{\'{e}}tien, I.~Arnal, and R.~H. Wade.
\newblock {Structural changes accompanying GTP hydrolysis in microtubules:
  information from a slowly hydrolyzable analogue
  guanylyl-(alpha,beta)-methylene-diphosphonate.}
\newblock {\em The Journal of cell biology}, 128(1-2):117--25, jan 1995.

\bibitem{Inoue2013}
D.~Inoue, A.~M.~R. Kabir, H.~Mayama, J.~P. Gong, K.~Sada, and A.~Kakugo.
\newblock {Growth of ring-shaped microtubule assemblies through stepwise active
  self-organisation}.
\newblock {\em Soft Matter}, 9(29):7061, jul 2013.

\bibitem{Inoue2015}
D.~Inoue, B.~Mahmot, A.~M.~R. Kabir, T.~I. Farhana, K.~Tokuraku, K.~Sada,
  A.~Konagaya, and A.~Kakugo.
\newblock {Depletion force induced collective motion of microtubules driven by
  kinesin.}
\newblock {\em Nanoscale}, 7(43):18054--61, oct 2015.

\bibitem{John2009a}
A.~John, A.~Schadschneider, D.~Chowdhury, and K.~Nishinari.
\newblock {Trafficlike collective movement of ants on trails: Absence of a
  jammed phase}.
\newblock {\em Physical Review Letters}, 102(10):108001, mar 2009.

\bibitem{Kabir2012}
A.~M.~R. Kabir, D.~Inoue, A.~Kakugo, K.~Sada, and J.~P. Gong.
\newblock {Active self-organization of microtubules in an inert chamber
  system}.
\newblock {\em Polymer Journal}, 44(6):607--611, apr 2012.

\bibitem{Kakugo2011}
A.~Kakugo, N.~Hosoda, K.~Shikinaka, and J.~P. Gong.
\newblock {Controlled Clockwise - Counterclockwise Motion of the Ring-Shaped
  Microtubules Assembly}.
\newblock {\em Biomacromolecules}, 12:3394--3399, 2011.

\bibitem{Katz2011}
Y.~Katz, K.~Tunstr{\o}m, C.~C. Ioannou, C.~Huepe, and I.~D. Couzin.
\newblock {Inferring the structure and dynamics of interactions in schooling
  fish.}
\newblock {\em Proceedings of the National Academy of Sciences of the United
  States of America}, 108(46):18720--5, nov 2011.

\bibitem{Kawaguchi2017}
K.~Kawaguchi, R.~Kageyama, and M.~Sano.
\newblock {Topological defects control collective dynamics in neural progenitor
  cell cultures}.
\newblock {\em Nature}, 545(7654):327--331, 2017.

\bibitem{Kawamura2008}
R.~Kawamura, A.~Kakugo, K.~Shikinaka, Y.~Osada, and J.~P. Gong.
\newblock {Ring-Shaped Assembly of Microtubules Shows Preferential
  Counterclockwise Motion}.
\newblock {\em Biomacromolecules}, 9(9):2277--2282, sep 2008.

\bibitem{Kim2018}
K.~Kim, N.~Yoshinaga, S.~Bhattacharyya, H.~Nakazawa, M.~Umetsu, and W.~Teizer.
\newblock {Soft Matter microtubules and kinesin motor proteins}.
\newblock {\em Soft Matter}, 14(17):3221--3231, may 2018.

\bibitem{Kraikivski2006}
P.~Kraikivski, R.~Lipowsky, and J.~Kierfeld.
\newblock {Enhanced ordering of interacting filaments by molecular motors}.
\newblock {\em Physical Review Letters}, 96(25):1--4, 2006.

\bibitem{Kruse2005a}
K.~Kruse, J.-F. Joanny, F.~Julicher, J.~Prost, and K.~Sekimoto.
\newblock {Generic theory of active polar gels: a paradigm for cytoskeletal
  dynamics}.
\newblock {\em European Physical Journal E}, 16(1):5--16, jun 2004.

\bibitem{Lam2014a}
A.~T. Lam, C.~Curschellas, D.~Krovvidi, and H.~Hess.
\newblock {Controlling self-assembly of microtubule spools via kinesin motor
  density}.
\newblock {\em Soft Matter}, 10(43):8731--8736, 2014.

\bibitem{Lauga2006}
E.~Lauga, W.~R. DiLuzio, G.~M. Whitesides, and H.~A. Stone.
\newblock {Swimming in circles: motion of bacteria near solid boundaries.}
\newblock {\em Biophysical journal}, 90(2):400--12, jan 2006.

\bibitem{Li1996}
J.-T. Li, J.~Carlsson, L.~Jinn-Nan, and K.~D. Caldwell.
\newblock {Chemical Modification of Surface Active Poly(ethylene
  oxide)-Poly(propylene oxide) Triblock Copolymers}.
\newblock {\em Bioconjugate Chemistry}, 7(5):592--599, 1996.

\bibitem{Liu2011b}
L.~Liu, E.~T{\"{u}}zel, and J.~L. Ross.
\newblock {Loop formation of microtubules during gliding at high density}.
\newblock {\em Journal of Physics Condensed Matter}, 23(37), 2011.

\bibitem{Lopez2012}
U.~Lopez, J.~Gautrais, I.~D. Couzin, and G.~Theraulaz.
\newblock {From behavioural analyses to models of collective motion in fish
  schools.}
\newblock {\em Interface focus}, 2(6):693--707, dec 2012.

\bibitem{Maeda1976}
K.~Maeda, Y.~Imae, J.~I. Shioi, and F.~Oosawa.
\newblock {Effect of temperature on motility and chemotaxis of Escherichia
  coli.}
\newblock {\em Journal of bacteriology}, 127(3):1039--46, sep 1976.

\bibitem{Samuel2012}
S.~R. McCandlish, A.~Baskaran, and M.~F. Hagan.
\newblock {Spontaneous segregation of self-propelled particles with different
  motilities}.
\newblock {\em Soft Matter}, 8(8):2527, 2012.

\bibitem{Narayan2007a}
V.~Narayan, S.~Ramaswamy, and N.~Menon.
\newblock {Long-lived giant number fluctuations in a swarming granular
  nematic}.
\newblock {\em Science}, 317(5834):105--108, 2007.

\bibitem{Nedelec1997}
F.~J. N{\'{e}}d{\'{e}}lec, T.~Surrey, a.~C. Maggs, and S.~Leibler.
\newblock {Self-organization of microtubules and motors.}
\newblock {\em Nature}, 389(6648):305--308, 1997.

\bibitem{Nishiguchi2018}
D.~Nishiguchi, J.~Iwasawa, H.~R. Jiang, and M.~Sano.
\newblock {Flagellar dynamics of chains of active Janus particles fueled by an
  AC electric field}.
\newblock {\em New Journal of Physics}, 20(1):015002, jan 2018.

\bibitem{Nishiguchi2017}
D.~Nishiguchi, K.~H. Nagai, H.~Chat{\'{e}}, and M.~Sano.
\newblock {Long-range nematic order and anomalous fluctuations in suspensions
  of swimming filamentous bacteria}.
\newblock {\em Physical Review E}, 95(2):020601, feb 2017.

\bibitem{Oswald2008}
P.~Oswald and A.~Dequidt.
\newblock {Measurement of the Continuous Lehmann Rotation of Cholesteric
  Droplets Subjected to a Temperature Gradient}.
\newblock {\em Physical Review Letters}, 100(21):217802, may 2008.

\bibitem{Peruani2017}
F.~Peruani.
\newblock {Hydrodynamic Equations for Flocking Models without Velocity
  Alignment}.
\newblock {\em Journal of the Physical Society of Japan}, 86(10):101010, oct
  2017.

\bibitem{Peruani2012}
F.~Peruani, J.~Starru{\ss}, V.~Jakovljevic, L.~S{\o}gaard-Andersen, A.~Deutsch,
  and M.~B{\"{a}}r.
\newblock {Collective Motion and Nonequilibrium Cluster Formation in Colonies
  of Gliding Bacteria}.
\newblock {\em Physical Review Letters}, 108(9):098102, feb 2012.

\bibitem{Ray1993}
S.~Ray, E.~Meyh{\"{o}}fer, R.~A. Milligan, and J.~Howard.
\newblock {Kinesin follows the microtubule's protofilament axis.}
\newblock {\em The Journal of cell biology}, 121(5):1083--93, jun 1993.

\bibitem{Saito2017}
A.~Saito, T.~I. Farhana, A.~M.~R. Kabir, D.~Inoue, A.~Konagaya, K.~Sada, and
  A.~Kakugo.
\newblock {Understanding the emergence of collective motion of microtubules
  driven by kinesins: role of concentration of microtubules and depletion
  force}.
\newblock {\em RSC Adv.}, 7(22):13191--13197, 2017.

\bibitem{Schaller2013}
V.~Schaller and A.~R. Bausch.
\newblock {Topological defects and density fluctuations in collectively moving
  systems}.
\newblock {\em Proceedings of the National Academy of Sciences},
  110(12):4488--4493, 2013.

\bibitem{Schaller2011a}
V.~Schaller, C.~Weber, E.~Frey, and A.~R. Bausch.
\newblock {Polar pattern formation: hydrodynamic coupling of driven filaments}.
\newblock {\em Soft Matter}, 7(7):3213, mar 2011.

\bibitem{Schaller2010}
V.~Schaller, C.~Weber, C.~Semmrich, E.~Frey, and A.~R. Bausch.
\newblock {Polar patterns of driven filaments.}
\newblock {\em Nature}, 467(7311):73--77, 2010.

\bibitem{Schaller2011}
V.~Schaller, C.~a. Weber, B.~Hammerich, E.~Frey, and A.~R. Bausch.
\newblock {Frozen steady states in active systems.}
\newblock {\em Proceedings of the National Academy of Sciences of the United
  States of America}, 108(48):19183--8, nov 2011.

\bibitem{Simha2002}
R.~A. Simha and S.~Ramaswamy.
\newblock {Hydrodynamic fluctuations and instabilities in ordered suspensions
  of self-propelled particles}.
\newblock {\em Physical review letters}, 89(5):058101, jul 2002.

\bibitem{Sumino2012}
Y.~Sumino, K.~H. Nagai, Y.~Shitaka, D.~Tanaka, K.~Yoshikawa, H.~Chat{\'{e}},
  and K.~Oiwa.
\newblock {Large-scale vortex lattice emerging from collectively moving
  microtubules}.
\newblock {\em Nature}, 483(7390):448--452, 2012.

\bibitem{Surrey2001}
T.~Surrey, F.~Nedelec, S.~Leibler, and E.~Karsenti.
\newblock {Physical properties determining self-organization of motors and
  microtubules.}
\newblock {\em Science}, 292(5519):1167--71, may 2001.

\bibitem{Suzuki2017}
R.~Suzuki and A.~R. Bausch.
\newblock {The emergence and transient behaviour of collective motion in active
  filament systems}.
\newblock {\em Nature Communications}, 8(1), 2017.

\bibitem{Szabo2006}
B.~Szab{\'{o}}, G.~J. Sz{\"{o}}ll{\"{o}}si, B.~G{\"{o}}nci, Z.~Jur{\'{a}}nyi,
  D.~Selmeczi, and T.~Vicsek.
\newblock {Phase transition in the collective migration of tissue cells:
  Experiment and model}.
\newblock {\em Physical Review E}, 74(6):061908, dec 2006.

\bibitem{Tamura2011}
Y.~Tamura, R.~Kawamura, K.~Shikinaka, A.~Kakugo, Y.~Osada, J.~P. Gong, and
  H.~Mayama.
\newblock {Dynamic self-organization and polymorphism of microtubule assembly
  through active interactions with kinesin}.
\newblock {\em Soft Matter}, 7(12):5654, jun 2011.

\bibitem{Tennenbaum2015}
M.~Tennenbaum, Z.~Liu, D.~Hu, and A.~Fernandez-Nieves.
\newblock {Mechanics of fire ant aggregations.}
\newblock {\em Nature materials}, 15, oct 2015.

\bibitem{Toner1995}
J.~Toner and Y.~Tu.
\newblock {Long-Range Order in a Two-Dimensional Dynamical XY Model: How Birds
  Fly Together}.
\newblock {\em Physical Review Letters}, 75(23):4326--4329, dec 1995.

\bibitem{Torisawa2016a}
T.~Torisawa, D.~Taniguchi, S.~Ishihara, and K.~Oiwa.
\newblock {Spontaneous Formation of a Globally Connected Contractile Network in
  a Microtubule-Motor System}.
\newblock {\em Biophysical Journal}, 111(2):373--385, jul 2016.

\bibitem{Vicsek1995}
T.~Vicsek, A.~Czir{\'{o}}k, E.~Ben-Jacob, I.~Cohen, and O.~Shochet.
\newblock {Novel type of phase transition in a system of self-driven
  particles.}
\newblock {\em Physical review letters}, 75(6):1226--1229, aug 1995.

\bibitem{Vicsek2012}
T.~Vicsek and A.~Zafeiris.
\newblock {Collective motion}.
\newblock {\em Physics Reports}, 517(3-4):71--140, 2012.

\bibitem{Wu2009}
Y.~Wu, A.~D. Kaiser, Y.~Jiang, and M.~S. Alber.
\newblock {Periodic reversal of direction allows Myxobacteria to swarm.}
\newblock {\em Proceedings of the National Academy of Sciences of the United
  States of America}, 106(4):1222--7, jan 2009.

\bibitem{Yamamoto2017}
T.~Yamamoto and M.~Sano.
\newblock {Chirality-induced helical self-propulsion of cholesteric liquid
  crystal droplets}.
\newblock {\em Soft Matter}, 13(18):3328--3333, may 2017.

\bibitem{Yamamoto2018}
T.~Yamamoto and M.~Sano.
\newblock {Theoretical model of chirality-induced helical self-propulsion}.
\newblock {\em Physical Review E}, 97(1):012607, jan 2018.

\bibitem{Yang2010}
Y.~Yang, V.~Marceau, and G.~Gompper.
\newblock {Swarm behavior of self-propelled rods and swimming flagella}.
\newblock {\em Physical Review E}, 82(3):031904, sep 2010.

\bibitem{Zhang2010}
H.~P. Zhang, A.~Be'er, E.-L. Florin, and H.~L. Swinney.
\newblock {Collective motion and density fluctuations in bacterial colonies.}
\newblock {\em Proceedings of the National Academy of Sciences of the United
  States of America}, 107(31):13626--30, aug 2010.

\bibitem{Ziebert2015}
F.~Ziebert, H.~Mohrbach, and I.~M. Kuli{\'{c}}.
\newblock {Why Microtubules Run in Circles: Mechanical Hysteresis of the
  Tubulin Lattice}.
\newblock {\em Physical Review Letters}, 114(14):148101, apr 2015.

\end{thebibliography}

\end{document}